\documentclass[epsfig,12pt,onecolumn]{article}
\usepackage{amsfonts}
\usepackage{amssymb}
\usepackage{multicol}
\usepackage{graphicx}
\usepackage{float}
\usepackage{caption}
\usepackage{xcolor}
\usepackage[utf8]{inputenc}
\usepackage{amsmath}

\makeatletter \@addtoreset{equation}{section}
\renewcommand{\theequation}{\thesection.\arabic{equation}}
\def \be{\begin{equation}}
\def \ee{\end{equation}}
\def \bea{\begin{eqnarray}}
\def \eea{\end{eqnarray}}

\newcommand{\nc}{\newcommand}
\nc{\al}{\alpha} \nc{\bib}{\bibitem} \nc{\la}{\lambda}
\nc{\C}{\mbox{\hspace{1.24mm}\rule{0.2mm}{2.5mm}\hspace{-2.7mm} C}}
\nc{\R}{\mbox{\hspace{.04mm}\rule{0.2mm}{2.8mm}\hspace{-1.5mm} R}}
\input{tcilatex}
\begin{document}

\title{\textbf{Algebraic constructions of code lattices in Narain CFTs}}
\author{E.H Saidi, R. Sammani \\
{\small 1. LPHE-MS, Faculty of Sciences, Mohammed V University in Rabat,
Morocco}\\
{\small 2. Hassan II Academy of Science and Technology, Kingdom of Morocco.}%
\\
{\small 3. Centre of Physics and Mathematics, CPM- Morocco.}}
\maketitle

\begin{abstract}
We give new results \textrm{on the structure and} representations of the
\textrm{three lattices }$\mathbf{\Lambda }_{\mathrm{k}},\mathbf{\Lambda }_{%
\mathrm{k}\mathcal{C}},\mathbf{\Lambda }_{\mathrm{k}}^{\ast }$ \textrm{%
relevant to} code CFTs realising Narain conformal field theories. In this
\textrm{construction,} $\mathbf{\Lambda }_{\mathrm{k}}^{\ast }$ \textrm{%
denotes} the dual of the even \textrm{lattice} $\mathbf{\Lambda }_{\mathrm{k}%
}$ and $\mathbf{\Lambda }_{\mathrm{k}\mathcal{C}}$ is an even self dual%
\textrm{\ intermediate lattice} with (d,d) signature. We \textrm{study} the
inclusion relations $\mathbf{\Lambda }_{\mathrm{k}}\subset \mathbf{\Lambda }%
_{\mathrm{k}\mathcal{C}}\subset \mathbf{\Lambda }_{\mathrm{k}}^{\ast }$
\textrm{characterised by} the discriminant \textrm{group} $\mathbf{\Lambda }%
_{\mathrm{k}}^{\ast }/\mathbf{\Lambda }_{\mathrm{k}}$ isomorphic to $\mathbb{%
Z}_{\mathrm{k}}$. \textrm{Explicitly constructions of these }$\mathbb{R}^{(%
\mathrm{r}d,\mathrm{r}d)}$\textrm{\ lattices are also provided }first for
rank $\mathrm{r}=d=1$ and then for higher dimensional Lie algebras with $%
\mathrm{r}=d>1$ . \textrm{Additional structural features and generalisations
are also discussed.}\newline
\textbf{Keywords}: Narain CFTs\ and coding theory, code lattices and self
duality condition, root /weight lattices of Lie algebras, construction A for
binary codes and extension.
\end{abstract}

\tableofcontents

\section{Introduction}

Recently, new connections have been established at the intersection of
quantum field theory, string compactifications, and quantum information
theory following the foundation of the bridge between Narain CFTs and
quantum error-correcting codes (QECs) \cite{Q1}-\cite{Q3}. The pivotal idea
is that the structure of Narain even self dual Lorentzian lattices,
parameterised by momenta and winding modes, can be naturally mapped to qubit
stabilizer codes, therefore extending Construction A of Euclidean lattices
from classical error correcting codes \cite{C1}-\cite{C3}. Hence, given the
emergence of Narain lattices in describing consistent error correcting
codes, understanding their construction, how they embed into larger dual
structure, how they relate through inclusion and discriminant groups is of
immense important for uncovering the algebraic properties of codes in
question. Accordingly, we aim in this work to provide concrete realisations
of these lattices in terms of finite dimensional Lie algebras as a way to
construct code models.\newline
Narain Conformal Field Theories (NCFTs) were originally developed in the
context of string theory, particularly in the study of toroidal
compactifications \cite{1A,1B}. These theories are characterized by a rich
symmetry structure and by the presence of a Narain lattice $\mathbf{\Lambda }%
_{\text{\textsc{narain}}}$ that encapsulates the spectrum generated by
Kaluza-Klein (KK) momentum states $|\vec{p}_{n_{KK}}>$ and string winding
modes $|\vec{p}_{m_{wind}}>$, thus providing a unified framework for
understanding the spectrum of closed strings in compactified spaces \cite{1}-%
\cite{4}. The Narain lattice is an even unimodular with signature (d,d)
embedded in $\mathbb{R}^{(d,d)}$ and coordinated by quantized left and right
momenta ($\vec{p}_{L},\vec{p}_{R}$). In the simplest case of
compactification on a circle $\mathbb{S}_{\text{\textsc{r}}}^{1}$ of radius
\textsc{r}, the KK momentum $p_{n_{KK}}$ along the compact direction is
quantized as $n/$\textsc{r} with $n\in \mathbb{Z}$ forming a 1D lattice $%
\mathbb{L}_{KK}=\mathtt{a}\mathbb{Z}$ (denoted just $\mathbb{L}$
subsequently for convenience) with parameter $\mathtt{a}=1$/\textsc{r}
assumed below like $\mathtt{a}\geq 1$ corresponding to a small radii \textsc{%
r}$\leq 1$. Regarding the winding modes, the $p_{m_{wind}}$ are also
quantized and given by $p_{m_{wind}}=m$\textsc{r}$.$ They coordinate a 1D
lattice $\mathbb{L}_{win}=\mathtt{b}\mathbb{Z=L}^{\ast }$ with parameter $%
\mathtt{b}=$\textsc{r} that can be\textrm{\ }thought of in terms of the
T-dual circle $\mathbb{S}_{\text{\textsc{\~{r}}}}^{1\ast }=\mathbb{S}_{1/%
\text{\textsc{r}}}^{1}$. Accordingly,\ the left $p_{L}$ and the right $p_{R}$
moving momenta of the particle states\ spanning\textrm{\ }$\dim \mathbf{%
\Lambda }_{\text{\textsc{narain}}}=2$\textrm{\ }are respectively given by $%
p_{\text{\textsc{l}/\textsc{r}}}=\frac{1}{2\text{\textsc{r}}}n\pm m$\textsc{r%
} \cite{5}; they take values in the 2D lattice $\mathbb{L}\times \mathbb{L}%
^{\ast }=\mathtt{a}\mathbb{Z}\times \mathtt{b}\mathbb{Z}$ that can
identified with the even lattice as follows
\begin{equation}
\mathbb{L}\times \mathbb{L}^{\ast }{\small \quad }\simeq {\small \quad }%
\mathbf{\Lambda }^{(1,1)}\qquad ,\qquad \mathbf{\Lambda }^{(1,1)}{\small %
\quad }\subset {\small \quad }\mathbb{R}\times \mathbb{R}  \label{1a}
\end{equation}%
where $\mathbf{\Lambda }^{(1,1)}\simeq \mathbf{\Lambda }_{\text{\textsc{%
narain}}}$ and where $\mathbb{R}^{(1,1)}$ is endowed with a hyperbolic
metric $\eta _{xy}$ ($\det \eta =-1$). In this setting, we have the
(duality)\ condition $\mathtt{a}.\mathtt{b}=1$ indicating that $\mathtt{%
b=1/a\leq 1}$ which leads to the inclusion property $\mathbb{L}\subset
\mathbb{L}^{\ast }.$ Notice that the lattice (\ref{1a}) has two closely
related constructions inspired from the left/right structures allowing to
think about $\mathbf{\Lambda }^{(1,1)}$ as part of the ensemble $\{\mathbf{%
\Lambda }^{(r,s)},$\quad $r+s=2\}.$ First, consider the lattice $\Lambda
^{(2,0)}:=\mathbb{L}\times \mathbb{L}$ which by substituting corresponds to
the square lattice \texttt{a}$\mathbb{Z}\times \mathtt{a}\mathbb{Z}.$ The
area of its unit cell denoted like $uc\mathbf{\Lambda }^{{\small (2,0)}}$ is
equal to $\mathtt{a}^{2}\geq 1$ since $\mathtt{a}\geq 1$. Second, we have
the lattice $\mathbf{\Lambda }^{(0,2)}:=\mathbb{L}^{\ast }\times \mathbb{L}%
^{\ast }$ which becomes $\mathtt{b}\mathbb{Z}\times \mathtt{b}\mathbb{Z}$
under substitution; the unit cell has area $uc\mathbf{\Lambda }^{{\small %
(0,2)}}=1{\small /}\mathtt{a}^{2}\leq 1$. This larger lattice is the dual of
$\mathbf{\Lambda }^{(2,0)}$ and contains $\mathbf{\Lambda }^{(1,1)}.$
Notably, the unit cell area of $\mathbb{L}\times \mathbb{L}^{\ast }$ is
equal to $uc\mathbf{\Lambda }^{{\small (1,1)}}=\mathtt{ab}=1$ due to the
self duality relation\textrm{\ }$\mathtt{b}=1/\mathtt{a}$. This
unimodularity serves as a criterion of self duality property of the Narain
lattice; see also \emph{Appendix A} for further details. As a result, the
inclusion relations of the triple of lattices take the form
\begin{equation}
\mathbf{\Lambda }^{(2,0)}{\small \quad }\subset {\small \quad }\mathbf{%
\Lambda }^{(1,1)}{\small \quad }\subset {\small \quad }\mathbf{\Lambda }%
^{(0,2)}\qquad \Leftrightarrow \qquad \mathbf{\Lambda }{\small \quad }%
\subset {\small \quad }\mathbf{\Lambda }_{\mathcal{C}}\subset {\small \quad }%
\mathbf{\Lambda }^{\ast }  \label{2a}
\end{equation}%
Here, the self dual $\mathbf{\Lambda }_{\mathcal{C}}$\ corresponds to the
Narain lattice\textrm{\ }$\mathbf{\Lambda }_{\text{\textsc{narain}}},$ and
the subscript $\mathcal{C}$ refers to "code" in the language of code CFT
\cite{1,2,5B}-\cite{5E}. In terms of the area of the unit cells, the
relations (\ref{2a}) translate into the inequalities
\begin{equation}
uc\mathbf{\Lambda }{\small \quad }>\quad 1\quad >{\small \quad }uc\mathbf{%
\Lambda }^{\ast }
\end{equation}%
where we used $uc\mathbf{\Lambda }_{\mathcal{C}}=1$. Observe also that in
the lattice $\mathbf{\Lambda }^{(1,1)},$ equipped with Lorentzian metric $%
\eta _{xy},$ the quantity $\mathcal{I}=p_{L}^{2}-p_{R}^{2},$ factorising
like $2h\times q$ with $h=\left( p_{L}+p_{R}\right) /\sqrt{2}$ and $q=\left(
p_{L}-p_{R}\right) /\sqrt{2},$ is independent of the radius \textsc{r} and
takes values in the set of even integers ($\mathcal{I}\in 2\mathbb{Z}$). For
higher dimensional even self-dual lattices $\mathbf{\Lambda }^{(d,d)}$ of
signature $(d,d),$ the moduli space's topological index $\mathcal{I}$
generalises to $\vec{p}_{L}^{2}-\vec{p}_{R}^{2}$ where the left/right
momenta ($\vec{p}_{n,m})_{\text{\textsc{l}/\textsc{r}}}$ have the structure $%
\sum_{i=1}^{d}(a_{n}^{i}\vec{\alpha}_{i}\pm b_{m}^{i}\vec{\lambda}_{i})/%
\sqrt{2}.$ Here, the basis vectors satisfy the duality condition $\vec{\alpha%
}_{i}.\vec{\lambda}^{j}=\delta _{i}^{j}$ and their mutual inner products are
encoded in the matrices $\vec{\alpha}_{i}.\vec{\alpha}_{j}=K_{ij}$ and $\vec{%
\lambda}^{i}.\vec{\lambda}^{j}=\tilde{K}^{ij}$. These momenta emerge
naturally in the framework of quantum error-correcting codes \cite{5}-\cite%
{7}, where the expansion coefficients in $\vec{p}_{\text{\textsc{l}/\textsc{r%
}}}$ are interpreted as equivalent classes $\bar{a}_{n}^{i}:=a^{i}+\mathrm{k}%
n^{i}$ and $\bar{b}_{m}^{i}:=b^{i}+\mathrm{k}m^{i}$ with $a^{i},b^{i}\in $ $%
\mathbb{Z}_{\mathrm{k}}=\left\{ 0,1,...,\mathrm{k}-1\right\} $ and $k$ is a
positive (prime) integer. From these expressions, several interesting
features can be deduced. In particular, the quantities $\vec{h}_{n}=(\vec{p}%
_{\text{\textsc{l}}}+\vec{p}_{\text{\textsc{r}}})/\sqrt{2}$ and $\vec{q}%
_{m}=(\vec{p}_{\text{\textsc{l}}}-\vec{p}_{\text{\textsc{r}}})/\sqrt{2}$
expand like
\begin{equation}
\vec{h}_{n}=\frac{1}{\sqrt{2}}\sum_{i=1}^{d}\bar{a}_{n}^{i}\vec{\alpha}%
_{i}\quad \in \mathbb{L}^{{\small (d)}}\qquad ,\qquad \vec{q}_{m}=\frac{1}{%
\sqrt{2}}\sum_{i=1}^{d}\bar{b}_{m}^{i}\vec{\lambda}_{i}\quad \in \mathbb{L}%
^{\ast {\small (d)}}  \label{plr}
\end{equation}%
and respectively belong to the lattice $\mathbb{L}^{{\small (d)}}$ and its
dual $\mathbb{L}^{\ast {\small (d)}}.$ As useful features for dealing with
the above relations, notice the following: $(\mathbf{i})$ the parameter $k$
is a positive definite (prime) integer ($\mathrm{k}\in \mathbb{N}^{\ast }$),
interpreted in \cite{1,8,9} as the level $\mathrm{k}$ of the Chern-Simons
theory with $U(1)^{d}\times U(1)^{d}$ gauge symmetry \cite{MW, Decoded, R4}
and field action,%
\begin{equation}
\mathcal{S}\sim \frac{\mathrm{k}}{4\pi }\dint\nolimits_{\mathcal{M}_{{\small %
3D}}}Tr\left( \mathcal{A}\wedge d\mathcal{B}+\mathcal{B}\wedge d\mathcal{A}%
\right) +\dint\nolimits_{\partial \mathcal{M}_{{\small 3D}}}Tr\left(
\mathcal{A}_{z}\mathcal{A}_{\bar{z}}+\mathcal{B}_{z}\mathcal{B}_{\bar{z}%
}\right)
\end{equation}%
$(\mathbf{ii})$ The quantities $a+\mathrm{k}n:=\bar{a}$ and $b+\mathrm{k}m:=%
\bar{b}$ define equivalence classes in the group $\mathbb{Z}_{\mathrm{k}%
}\simeq \mathbb{Z}/(\mathrm{k}\mathbb{Z})$. These classes can be
alternatively imagined in terms of shifted fractions as $\bar{a}\sim \frac{a%
}{\mathrm{k}}+n$ or equivalently in terms of k-th roots of unity like $%
e^{i2\pi \frac{a}{\mathrm{k}}}.$ For the case k=1, these momenta reduce to
the standard values in Narain theory, as we have
\begin{equation}
k=1:\qquad \bar{a}\sim \bar{0}\qquad ,\qquad \bar{b}\sim \bar{0}
\end{equation}%
$(\mathbf{iii})$ Thinking of the $\vec{\alpha}_{i}$ \textrm{as} (d+d)
vectors $\vec{\alpha}_{i}^{x}=(\vec{\alpha}_{i},\vec{0})$ with label $x=1,2$%
, and about the $\vec{\lambda}_{i}$'s like $\vec{\lambda}^{xj}=(\vec{0},\vec{%
\lambda}_{i}),$ the signature (d,d) of the even self-dual lattice $\mathbf{%
\Lambda }^{(d,d)}$ is given by the bilinear pairing $\left\langle \mathbf{%
\alpha },\mathbf{\lambda }\right\rangle =\eta _{xy}\vec{\alpha}^{x}.\vec{%
\lambda}^{y}$ coinciding with the duality $\vec{\alpha}_{i}.\vec{\lambda}%
^{j}=\delta _{i}^{j}.$ Here the metric $\eta _{xy}$ is taken as in (\ref{26}%
).

The purpose of the paper is twofold: First, we study the structural
properties of the triplet $(\mathbf{\Lambda }_{\mathrm{k}},\mathbf{\Lambda }%
_{\mathrm{k}\mathcal{C}},\mathbf{\Lambda }_{\mathrm{k}}^{\ast })$\ arising
from code constructions of Narain CFT. This includes understanding the
relations between these lattices via their discriminant groups $\mathbf{%
\Lambda }^{\ast }/\mathbf{\Lambda }$, $\mathbf{\Lambda }^{\ast }/\mathbf{%
\Lambda }_{\mathcal{C}}$ and $\mathbf{\Lambda }_{\mathcal{C}}/\mathbf{%
\Lambda }$, examining their embeddings given by the inclusion relations $%
\mathbf{\Lambda }\subset \mathbf{\Lambda }_{\mathcal{C}}$ and $\mathbf{%
\Lambda }_{\mathcal{C}}\subset \mathbf{\Lambda }^{\ast }$ as well as
encoding new superposition structure of isomorphic sublattices that emerges
in the non orthogonal case where the intersection matrices $\mathbf{\alpha }%
_{i}.\mathbf{\alpha }_{j}$ and $\mathbf{\lambda }^{i}.\mathbf{\lambda }^{j}$
are non vanishing. Second, we illustrate our findings through concrete
examples by giving explicit low rank constructions based on the Lie algebras
su(2) and su(3). Our objective is not to classify all possible code-based
Narain lattices, but rather to study representative families as a bottom up
demonstration of how code CFTs emerge from the underlying lattice and
discriminant group data. Here below, we provide a concise summary of our
main results:

(i)- We construct concrete realisations of $\mathbf{\Lambda }_{\mathcal{C}}$
within $\mathbb{R}^{(d,d)}$\textrm{\ }and examine the structural aspects
like the multiplicities ${\small [}\mathbf{\Lambda }_{\mathcal{C}}{\small ]}%
_{\mathrm{j=1,...}}$ of the even self dual lattice in $\mathbb{R}^{(d,d)}$
and their mutual isomorphic relations. Because, in this study, the coset $%
\mathbf{\Lambda }^{\ast }/\mathbf{\Lambda }$ is isomorphic to a non trivial
finite group $G_{\mathrm{k}}\sim \mathbb{Z}_{\mathrm{k}}^{2\mathrm{d}}$ with
several elements as \{$I_{id},g_{2},...,g_{N_{\mathrm{k}}}$\}, we show
amongst others that the inclusion relations (\ref{2a}) should read in
general like

\begin{equation}
\begin{tabular}{ccccc}
&  & ${\small [}\mathbf{\Lambda }_{\mathcal{C}}{\small ]}_{\mathrm{1}}$ &  &
\\
& $\nearrow $ & $\vdots $ & $\searrow $ &  \\
$\mathbf{\Lambda }$ & $\longrightarrow $ & $\vdots $ & $\longrightarrow $ & $%
\mathbf{\Lambda }^{\ast }$ \\
& $\searrow $ & $\vdots $ & $\nearrow $ &  \\
&  & ${\small [}\mathbf{\Lambda }_{\mathcal{C}}{\small ]}_{M_{\mathrm{k}}}$
&  &
\end{tabular}
\label{3a}
\end{equation}%
\begin{equation*}
\end{equation*}%
In this formal diagram, the even self dual lattices ${\small [}\mathbf{%
\Lambda }_{\mathcal{C}}{\small ]}_{\mathrm{j}}$ are mutually isomorphic,
i.e: ${\small [}\mathbf{\Lambda }_{\mathcal{C}}{\small ]}_{\mathrm{1}}\simeq
{\small [}\mathbf{\Lambda }_{\mathcal{C}}{\small ]}_{\mathrm{2}}\simeq
...\simeq {\small [}\mathbf{\Lambda }_{\mathcal{C}}{\small ]}_{M_{\mathrm{k}%
}}$, and together form a multiplet of the discriminant group $G_{\mathrm{k}}$%
. We also give explicit realisations of the triples ($\mathbf{\Lambda },%
{\small [}\mathbf{\Lambda }_{\mathcal{C}}{\small ]}_{\mathrm{j}},\mathbf{%
\Lambda }^{\ast }$) sitting in $\mathbb{R}^{(d,d)}$ with $d\geq 1.$

(ii) We obtain new results on the representation of the embedding relations (%
\ref{1a}-\ref{3a}) in 2D and their extensions to higher dimensional
settings. Our construction reveals a natural distinction between two main
classes depending on the structure of the intersection matrices $\mathbf{%
\alpha }_{i}.\mathbf{\alpha }_{j}$ and $\mathbf{\lambda }^{i}.\mathbf{%
\lambda }^{j}$ of the generators $\mathbb{L}^{{\small (d)}}$ and its dual $%
\mathbb{L}^{\ast {\small (d)}}$: $\left( a\right) $ \emph{the orthogonal
case }with vanishing inner product (i.e: $\mathbf{\alpha }_{i}\perp \mathbf{%
\alpha }_{j}$ and $\mathbf{\lambda }^{i}\perp \mathbf{\lambda }^{j}$)
corresponding to the standard construction A of code CFT; and $\left(
b\right) $ \emph{the non-orthogonal case} developed here using lattices of
finite dimensional Lie algebras like the special su(r+1) family labeled by
the rank r determining the central charge of the associated Narain CFT.%
\newline
We also note that the present construction may be viewed as an alternative
formulation of the so called construction A$_{\mathbf{g}}$ investigated in
\textrm{\cite{71}; }\emph{see appendix B for a brief review}.

While completing this work, we became aware of the recent preprint by
Angelinos \cite{Ang}, which also considers even lattices constructed from
root lattices of simple Lie algebras. That work is restricted to level k=1,
whereas in the present paper we develop a structural analysis of the full
triplet $(\mathbf{\Lambda }_{\mathrm{k}},\mathbf{\Lambda }_{\mathrm{k}%
\mathcal{C}},\mathbf{\Lambda }_{\mathrm{k}}^{\ast })$ for arbitrary
Chern--Simons level k, including the discriminant decomposition, the
superposition of isomorphic sublattices, and the resulting multiplet
structure. These aspects, together with the explicit low-dimensional
realisations of the inclusion chain $\mathbf{\Lambda }_{\mathrm{k}}\subset
\mathbf{\Lambda }_{\mathrm{k}\mathcal{C}}\subset \mathbf{\Lambda }_{\mathrm{k%
}}^{\ast }$, are not addressed in \cite{Ang}.

The organisation is as follows: In section 2, we start by considering the
standard case of the two dimensional lattices ($\mathbf{\Lambda },\mathbf{%
\Lambda }_{\mathcal{C}},\mathbf{\Lambda }^{\ast }$) with structures as in (%
\ref{2a}). We first show that these three lattices are intimately related
with the root $\mathbf{\Lambda }_{R}^{\mathbf{su}_{2}}:=$R$^{\mathbf{su}%
_{2}} $ and the weight $\mathbf{\Lambda }_{W}^{\mathbf{su}_{2}}:=$W$^{%
\mathbf{su}_{2}}$ lattices of the su(2) Lie algebra. Then, we use the
obtained results to revisit the construction A of code theory associating
quantum stabilizer codes to even self-dual Lorentzian lattices thought of in
our description in terms of tensor products of R$^{\mathbf{su}_{2}}$ and W$^{%
\mathbf{su}_{2}}$. In section 3, we extend our su(2) based construction to
su(3) by using the root R$^{\mathbf{su}_{3}}$ and weight W$^{\mathbf{su}_{3}}
$ lattices. For this generalisation, we consider three situations according
to the value of the CS level \textrm{k}: the special value \textrm{k}=3, the
range \textrm{k}\TEXTsymbol{>}3 and the situations with k\TEXTsymbol{<}3.
This su(3) based description should be viewed as giving an illustration of
general buildings using root R$^{\mathbf{g}}$ and weight W$^{\mathbf{g}}$
lattices of finite Lie algebras \textbf{g}. In section 4, we give a
conclusion and make comments regarding the su(n) family. Useful tools and
technical details are moved to the appendix.

\section{$\mathbf{\Lambda }_{\mathrm{k}}^{\ast },$ $\mathbf{\Lambda _{%
\mathrm{k}}}$, $\mathbf{\Lambda }_{\mathrm{k}\mathcal{C}}$ lattices and
higher dimensional extensions}

We begin by examining the structures of the 2D lattices $\mathbf{\Lambda }_{%
\mathrm{k}}^{\ast },$ $\mathbf{\Lambda }_{\mathrm{k}}$ and $\mathbf{\Lambda }%
_{\mathrm{k}\mathcal{C}}$ labeled by the CS level $\mathrm{k.}$ We focus on
the particular case k=2 as\textrm{\ }this value recovers the exact root and
weight lattices of su(2). For this level, the chain of inclusions (\ref{2a}%
)\ reads
\begin{equation*}
\mathbf{\Lambda }_{\mathrm{2}}\quad \subset \quad \mathbf{\Lambda }_{\mathrm{%
2}\mathcal{C}}\quad \subset \quad \mathbf{\Lambda }_{\mathrm{2}}^{\ast
}\qquad ,\qquad \mathrm{k}=2
\end{equation*}%
We then consider generalised models with arbitrary prime levels $\mathrm{k}%
\in \mathbb{N}^{\ast }$, maintaining $\mathbf{\Lambda }_{\mathrm{k}}\subset
\mathbf{\Lambda }_{\mathrm{k}\mathcal{C}}\subset \mathbf{\Lambda }_{\mathrm{k%
}}^{\ast }$. Next, we extend the analysis to higher dimensional lattices
using construction A from codding theory. The analogue of the chain of
inclusions for lattices of dimension $\dim \mathbf{\Lambda }_{\mathrm{k}}>2$
can be presented like $\mathbf{\Lambda }_{\mathrm{k}}^{(2d,0)}\subset
\mathbf{\Lambda }_{\mathrm{k}}^{(d,d)}\subset \mathbf{\Lambda }_{\mathrm{k}%
}^{(0,2d)}$ with positive integer $d\geq 2$ and self dual\textrm{\ }$\mathbf{%
\Lambda }_{\mathrm{k}}^{(d,d)}$.

\subsection{Two dimensional lattices $\mathbf{\Lambda }_{\mathrm{2}}^{\ast
}, $ $\mathbf{\Lambda _{\mathrm{2}}}$, $\mathbf{\Lambda }_{\mathrm{2}%
\mathcal{C}},$ $\mathbf{\tilde{\Lambda}}_{\mathrm{2}\mathcal{C}}$}

\begin{figure}[tbph]
\begin{center}
\includegraphics[width=12cm]{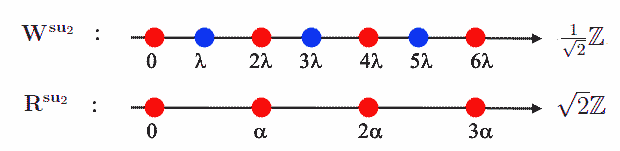}
\end{center}
\par
\vspace{-0.5cm}
\caption{At the top, we show the weight lattice $\mathbf{W}^{\mathbf{su}%
_{2}} $ of su(2)\textrm{\ }which is isomorphic to $\frac{1}{\protect\sqrt{2}}%
\mathbb{Z}$. At the bottom, we present the root lattice \ $\mathbf{R}^{%
\mathbf{su}_{2}}$ isomorphic to $\protect\sqrt{2}\mathbb{Z}.$ The\textrm{\ }%
quotient W$^{\mathbf{su}_{2}}/$R$^{\mathbf{su}_{2}}$ defines the
discriminant group, isomorphic to $\mathbb{Z}_{2}.\ $The lattices sites
\textbf{x}$_{n}$ are occupied by particles that can be described in terms of
wave functions $\protect\psi \left( \mathbf{x}_{n}\right) $.}
\label{a0}
\end{figure}

We start by introducing the weight W$^{\mathbf{su}_{2}}$ and the root R$^{%
\mathbf{su}_{2}}$ lattices of the su(2) Lie algebra. These are one
dimensional lattices generated, respectively, by the fundamental weight
vector $\mathbf{\lambda }$ and the simple root $\mathbf{\alpha }$ of su(2)
satisfying the standard relations:%
\begin{equation}
\mathbf{\lambda }.\mathbf{\lambda }=\frac{1}{2}\qquad ,\qquad \mathbf{%
\lambda }.\mathbf{\alpha }=1\qquad ,\qquad \mathbf{\alpha }.\mathbf{\alpha }%
=2
\end{equation}%
from which it follows that $\mathbf{\alpha }=2\mathbf{\lambda }$. In terms
of the canonical basis $\left\{ \mathbf{e}_{1},\mathbf{e}_{2}\right\} $ of $%
\mathbb{R}^{2}$ with euclidian metric, these vectors can be imagined as $%
\mathbf{\alpha =}\sqrt{2}\mathbf{u}$ and $\mathbf{\lambda }=\mathbf{u}/\sqrt{%
2}$ where $\mathbf{u}=\left( \mathbf{e}_{1}-\mathbf{e}_{2}\right) /\sqrt{2}.$
A graphical representation of these dual lattices is given in Figure \textbf{%
\ref{a0} }where red and blue dots indicate the lattices sites of R$^{\mathbf{%
su}_{2}}$ and W$^{\mathbf{su}_{2}}$ respectively\textrm{; }note that R$^{%
\mathbf{su}_{2}}\subset $W$^{\mathbf{su}_{2}}.$ These lattices are
isomorphic to $\frac{1}{\sqrt{2}}\mathbb{Z}$ and $\sqrt{2}\mathbb{Z}$ with
unit cell lengths $\left\vert \mathbf{\lambda }\right\vert =1/\sqrt{2}$ and $%
\left\vert \mathbf{\alpha }\right\vert =\sqrt{2}.$
The weight lattice W$^{\mathbf{su}_{2}}$\textrm{\ }has sites spaced at
half-integer intervals, and it contains the root sublattice R$^{\mathbf{su}%
_{2}}$ whose sites lie at even integer since $\mathbf{\alpha }^{2}=2$. The
inclusion property is captured by the discriminant group W$^{\mathbf{su}%
_{2}}/$R$^{\mathbf{su}_{2}}$ which is isomorphic to the group $\mathbb{Z}/2%
\mathbb{Z\simeq Z}_{2}.$ This indicates that W$^{\mathbf{su}_{2}}$ is the
superposition of two root sublattices $R^{\mathbf{su}_{2}}\cup \tilde{R}^{%
\mathbf{su}_{2}}$ where $\tilde{R}^{\mathbf{su}_{2}}$ (blue sites in Figure
\textbf{\ref{a0}} ) is shifted from $R^{\mathbf{su}_{2}}$ (red sites in
Figure \textbf{\ref{a0}}) by a fundamental weight vector $\mathbf{\lambda ;}$
that is $\tilde{R}^{\mathbf{su}_{2}}\simeq R^{\mathbf{su}_{2}}+\mathbf{%
\lambda }.$ For convenience, we set below $R^{\mathbf{su}_{2}}=\mathbb{A}$
and $\tilde{R}^{\mathbf{su}_{2}}=\mathbb{B}$; so we have W$^{\mathbf{su}%
_{2}}=\mathbb{A}\cup \mathbb{B}$ with the isomorphism $\mathbb{B}\simeq
\mathbb{A}+\mathbf{\lambda }.$\newline
The vector sites in W$^{\mathbf{su}_{2}}$ are of the form $\mathbf{w}=m%
\mathbf{\lambda }$ with m $\in \mathbb{Z}$ and have euclidian length $%
\mathbf{w}^{2}=\frac{1}{2}m^{2}\in \frac{1}{2}\mathbb{N}.$ While those in R$%
^{\mathbf{su}_{2}}$ are given by $\mathbf{r}=n\mathbf{\alpha }$ with n$\in
\mathbb{Z}$ and euclidian length $\mathbf{r}^{2}=2n^{2}\in 2\mathbb{N}.$
Additionally, the\textrm{\ }site positions \textbf{x}$_{n}$ in the lattices
of the Figure \textbf{\ref{a0}} are occupied by particle states interpreted
as hybrids of KK modes and windings with momenta like $p_{\text{\textsc{l}/%
\textsc{r}}}=\frac{1}{2\text{\textsc{r}}}n\pm m$\textsc{r} \cite{9}.
Thinking of these quantum states in terms of wave functions $\psi \left(
\mathbf{x}_{n}\right) ,$ allows further investigation into their coupling
properties via tight binding modeling akin to those in condensed matter
system \cite{CM1}-\cite{CM4}. The effective Hamiltonian in such models
typically takes the quadratic form \cite{DS1,DS2,DS3},\textrm{\ }%
\begin{equation}
H=\sum_{\text{lattice}}\psi _{\mathbf{x}_{\mathbf{n}}}t_{\mathbf{n},\mathbf{m%
}}\psi _{\mathbf{x}_{\mathbf{m}}}^{\dagger }+hc
\end{equation}%
where $t_{\mathbf{n},\mathbf{m}}$ encode the coupling between states at site
$\mathbf{x}_{\mathbf{n}}$ and site $\mathbf{x}_{\mathbf{m}}.$ Further
analysis of this Hamiltonian however lies beyond the scope of the present
paper.

\subsubsection{Lattice of $SU(2)\times SU(2)$}

Given the weight W$^{\mathbf{su}_{2}}$ and root R$^{\mathbf{su}_{2}}$
lattices of the Lie algebra su(2), one can construct higher dimensional
lattices through tensor products. In particular, we focus on the following
2D lattices:%
\begin{equation}
\begin{tabular}{|lll||cllll|}
\hline
\multicolumn{3}{|c||}{lattice} & \multicolumn{5}{|c|}{coordinates} \\
\hline\hline
$\mathbf{\Lambda }_{\mathrm{2}}^{\ast }$ & $=$ & $W^{\mathbf{su}_{2}}\times
W^{\prime \mathbf{su}_{2}}$ & $\mathbf{w}$ & $=$ & $m\mathbf{\lambda }%
+m^{\prime }\mathbf{\lambda }^{\prime }$ & $=$ & $\frac{1}{\sqrt{2}}\left( m%
\mathbf{u}+m^{\prime }\mathbf{u}^{\prime }\right) $ \\
$\mathbf{\Lambda }_{\mathrm{2}}$ & $=$ & $R^{\mathbf{su}_{2}}\times
R^{\prime \mathbf{su}_{2}}$ & \multicolumn{1}{||l}{$\mathbf{r}$} & $=$ & $n%
\mathbf{\alpha }+n^{\prime }\mathbf{\alpha }^{\prime }$ & $=$ & $\sqrt{2}%
\left( n\mathbf{u}+n^{\prime }\mathbf{u}^{\prime }\right) $ \\
$\mathbf{\Lambda }_{\mathrm{2}\mathcal{C}}$ & $=$ & $R^{\mathbf{su}%
_{2}}\times W^{\prime \mathbf{su}_{2}}$ & $\mathbf{x}$ & $=$ & $n\mathbf{%
\alpha }+m^{\prime }\mathbf{\lambda }^{\prime }$ & $=$ & $\sqrt{2}n\mathbf{u}%
+\frac{1}{\sqrt{2}}m^{\prime }\mathbf{u}^{\prime }$ \\
$\mathbf{\tilde{\Lambda}}_{\mathrm{2}\mathcal{C}}$ & $=$ & $W^{\mathbf{su}%
_{2}}\times R^{\prime \mathbf{su}_{2}}$ & \multicolumn{1}{||l}{$\mathbf{%
\tilde{x}}$} & $=$ & $m\mathbf{\lambda }+n^{\prime }\mathbf{\alpha }^{\prime
}$ & $=$ & $\frac{1}{\sqrt{2}}m\mathbf{u}+\sqrt{2}n^{\prime }\mathbf{u}%
^{\prime }$ \\ \hline\hline
\end{tabular}
\label{L}
\end{equation}%
\begin{equation*}
\end{equation*}%
Here\textrm{\ }$m,m^{\prime },n,n^{\prime }$\textrm{\ }are integers and the\
inner product satisfy $\mathbf{\lambda }.\mathbf{\lambda }^{\prime }=0=%
\mathbf{\alpha }.\mathbf{\alpha }^{\prime }$ as well as $\mathbf{\lambda }.%
\mathbf{\alpha }^{\prime }=\mathbf{\alpha }.\mathbf{\lambda }^{\prime }=0$.
Moreover, because W$^{\mathbf{su}_{2}}/$R$^{\mathbf{su}_{2}}\simeq \mathbb{Z}%
_{2},$ we obtain
\begin{equation}
\mathbf{\Lambda }_{\mathrm{2}}^{\ast }/\mathbf{\Lambda }_{\mathrm{2}}\simeq
\mathbb{Z}_{2}\times \mathbb{Z}_{2}\qquad ,\qquad \mathbf{\Lambda }_{\mathrm{%
2}}^{\ast }/\mathbf{\Lambda }_{2\mathcal{C}}\simeq \mathbb{Z}_{2}\qquad
,\qquad \mathbf{\Lambda }_{2\mathcal{C}}/\mathbf{\Lambda }_{\mathrm{2}%
}\simeq \mathbb{Z}_{2}
\end{equation}%
The corresponding unit cell areas are given by%
\begin{equation}
\begin{tabular}{|ccc||ccc|}
\hline
\multicolumn{3}{|c||}{unit cell: uc} & \multicolumn{3}{||c|}{area of unit
cell /$\mathbf{u}\wedge \mathbf{u}^{\prime }$} \\ \hline\hline
$\ \ \mathbf{\lambda }\wedge \mathbf{\lambda }^{\prime }$ & $=$ & $\frac{1}{2%
}\mathbf{u}\wedge \mathbf{u}^{\prime }$ \ \ \  & $\ \ \ \mathbf{uc(\mathbf{%
\Lambda }}_{\mathrm{2}}^{\ast }\mathbf{)}$ & $=$ & $\frac{1}{2}$ \\
$\ \ \mathbf{\alpha }\wedge \mathbf{\alpha }^{\prime }$ & $=$ & $2\mathbf{u}%
\wedge \mathbf{u}^{\prime }$ \ \ \ \  & $\ \ \ \mathbf{uc(\mathbf{\Lambda }}%
_{\mathrm{2}}\mathbf{)}$ & $=$ & $2$ \\
$\ \mathbf{\alpha }\wedge \mathbf{\lambda }^{\prime }$ & $=$ & $\mathbf{u}%
\wedge \mathbf{u}^{\prime }$ \ \ \ \  & $\ \ \ \mathbf{uc(\mathbf{\Lambda }}%
_{2\mathcal{C}}\mathbf{)}$ & $=$ & $1$ \\
$\ \ \mathbf{\lambda }\wedge \mathbf{\alpha }^{\prime }$ & $=$ & $-\mathbf{u}%
\wedge \mathbf{u}^{\prime }$ \ \ \ \ \  & $\ \ \ \mathbf{uc(\mathbf{\tilde{%
\Lambda}}}_{2\mathcal{C}}\mathbf{)}$ & $=$ & $1$ \\ \hline\hline
\end{tabular}%
\end{equation}%
\begin{equation*}
\end{equation*}%
For convenience, we may write these lattices as $\mathbf{\Lambda }_{\mathrm{2%
}}^{\ast }=\mathbb{Z}\mathbf{\lambda }\oplus \mathbb{Z}\mathbf{\lambda }%
^{\prime }$\ and $\mathbf{\Lambda }_{\mathrm{2}}=\mathbb{Z}\mathbf{\alpha }%
\oplus \mathbb{Z}\mathbf{\alpha }^{\prime }$\ as well as $\mathbf{\Lambda }_{%
\mathrm{2}\mathcal{C}}=\mathbb{Z}\mathbf{\alpha }\oplus \mathbb{Z}\mathbf{%
\lambda }^{\prime }$ and $\mathbf{\tilde{\Lambda}}_{\mathrm{2}\mathcal{C}}=%
\mathbb{Z}\mathbf{\lambda }\oplus \mathbb{Z}\mathbf{\alpha }^{\prime }.$ In
terms of the unit vectors $(\mathbf{u},\mathbf{u}^{\prime })$ which we label
as $(\mathbf{u}_{1},\mathbf{u}_{2})$, the lattices of eq(\ref{L}) take the
explicit form:
\begin{equation}
\begin{tabular}{lllll}
$\mathbf{\Lambda }_{\mathrm{2}}^{\ast }$ & $=$ & $\frac{1}{\sqrt{2}}\mathbb{Z%
}$ $\mathbf{u}_{1}$ & $\oplus $ & $\frac{1}{\sqrt{2}}\mathbb{Z}$ $\mathbf{u}%
_{2}$ \\
$\mathbf{\Lambda }_{\mathrm{2}}$ & $=$ & $\sqrt{2}\mathbb{Z}$ $\mathbf{u}%
_{1} $ & $\oplus $ & $\sqrt{2}\mathbb{Z}$ $\mathbf{u}_{2}$ \\
$\mathbf{\Lambda }_{\mathrm{2}\mathcal{C}}$ & $=$ & $\sqrt{2}\mathbb{Z}$ $%
\mathbf{u}_{1}$ & $\oplus $ & $\frac{1}{\sqrt{2}}\mathbb{Z}$ $\mathbf{u}_{2}$
\\
$\mathbf{\tilde{\Lambda}}_{\mathrm{2}\mathcal{C}}$ & $=$ & $\frac{1}{\sqrt{2}%
}\mathbb{Z}$ $\mathbf{u}_{1}$ & $\oplus $ & $\sqrt{2}\mathbb{Z}$ $\mathbf{u}%
_{2}$%
\end{tabular}%
\end{equation}%
Also, using the components $\mathbf{u}_{i}=\left( u_{i}^{\dot{a}}\right) $,
the metric $g_{ij}=\mathbf{u}_{i}.\mathbf{u}_{j}=u_{i}^{\dot{a}}\delta _{%
\dot{a}\dot{b}}u_{j}^{\dot{b}}$ can be defined through either the Euclidian $%
\delta _{ij}$ or the Lorentzian $\eta _{ij}$ given in terms of the 2$\times $%
2 matrices
\begin{equation}
\delta _{ij}=\left(
\begin{array}{cc}
1 & 0 \\
0 & 1%
\end{array}%
\right) \qquad ,\qquad \eta _{ij}=\left(
\begin{array}{cc}
0 & 1 \\
1 & 0%
\end{array}%
\right)  \label{26}
\end{equation}%
implying that $g_{ij}^{\left( \text{\textsc{e}}\right) }=\left\langle
\mathbf{u}_{i}.\mathbf{u}_{j}\right\rangle _{\text{\textsc{e}}}=\delta _{ij}$
for an euclidian geometry and $g_{ij}^{\left( \text{\textsc{l}}\right)
}=\left\langle \mathbf{u}_{i}.\mathbf{u}_{j}\right\rangle _{\text{\textsc{l}}%
}=\eta _{ij}$, in a Lorentzian case. Accordingly, for two site vectors in
the lattices (\ref{L}) $\mathbf{x}=x^{1}\mathbf{u}_{1}+x^{2}\mathbf{u}_{2}$
and $\mathbf{y}=y^{1}\mathbf{u}_{1}+y^{2}\mathbf{u}_{2},$ the
(Euclidian/Lorentzian) pairing are given by%
\begin{equation}
\begin{tabular}{lllll}
$\left\langle \mathbf{x}.\mathbf{y}\right\rangle _{\text{\textsc{e}}}$ & $=$
& $x^{i}\delta _{ij}y^{j}$ & $=$ & $x^{1}y^{1}+x^{2}y^{2}$ \\
$\left\langle \mathbf{x}.\mathbf{y}\right\rangle _{\text{\textsc{l}}}$ & $=$
& $x^{i}\eta _{ij}y^{j}$ & $=$ & $x^{1}y^{2}+x^{2}y^{1}$%
\end{tabular}%
\end{equation}%
For the case where $\mathbf{y}=\mathbf{x}$, they reduce to $\left\langle
\mathbf{x}.\mathbf{x}\right\rangle _{\text{\textsc{e}}}=\left( x^{1}\right)
^{2}+\left( x^{2}\right) ^{2}=\left( x_{1}\right) ^{2}+\left( x_{2}\right)
^{2}$ and $\left\langle \mathbf{x}.\mathbf{x}\right\rangle _{\text{\textsc{l}%
}}=2x^{1}x^{2}=2x_{1}x_{2}.$ Per illustration, we have:%
\begin{equation}
\begin{tabular}{|c|ccc|c|c|c|}
\hline
{\small lattice} & \multicolumn{3}{|c|}{\small site vectors} & $\left\langle
\mathbf{\ast }.\mathbf{\ast }\right\rangle _{\text{\textsc{e}}}$ & $%
\left\langle \mathbf{\ast }.\mathbf{\ast }\right\rangle _{\text{\textsc{l}}}$
& {\small unit cell} \\ \hline
$\mathbf{\Lambda }_{\mathrm{2}}^{\ast }$ & $\mathbf{w}$ & $=$ & $\frac{1}{%
\sqrt{2}}\left( m^{1}\mathbf{u}_{1}+m^{2}\mathbf{u}_{2}\right) $ & $\frac{1}{%
2}\left( m_{1}\right) ^{2}+\frac{1}{2}\left( m_{2}\right) ^{2}$ & $%
m_{2}m_{1} $ & $\frac{1}{2}$ \\
$\mathbf{\Lambda }_{\mathrm{2}}$ & $\mathbf{r}$ & $=$ & $\sqrt{2}\left( n^{1}%
\mathbf{u}_{1}+n^{2}\mathbf{u}_{2}\right) $ & $2\left( n_{1}\right)
^{2}+2\left( n_{2}\right) ^{2}$ & $4n_{2}n_{1}$ & $2$ \\
$\mathbf{\Lambda }_{\mathrm{2}\mathcal{C}}$ & $\mathbf{x}$ & $=$ & $\sqrt{2}%
n^{1}\mathbf{u}_{1}+\frac{1}{\sqrt{2}}m^{2}\mathbf{u}_{2}$ & $2\left(
n_{1}\right) ^{2}+\frac{1}{2}\left( m_{2}\right) ^{2}$ & $2n_{2}m_{1}$ & $1$
\\
$\mathbf{\tilde{\Lambda}}_{\mathrm{2}\mathcal{C}}$ & $\mathbf{\tilde{x}}$ & $%
=$ & $\frac{1}{\sqrt{2}}m^{1}\mathbf{u}_{1}+\sqrt{2}n^{2}\mathbf{u}_{2}$ & $%
\frac{1}{2}\left( m_{1}\right) ^{2}+2\left( n_{2}\right) ^{2}$ & $%
2m_{2}n_{1} $ & $1$ \\ \hline
\end{tabular}%
\end{equation}

\subsubsection{Properties of $(\mathbf{\Lambda _{\mathrm{2}},\Lambda }_{%
\mathrm{2}\mathcal{C}},\mathbf{\tilde{\Lambda}}_{\mathrm{2}\mathcal{C}},%
\mathbf{\Lambda }_{\mathrm{2}}^{\ast })$}

Based on the preceding construction, various properties of the 2D lattices $(%
\mathbf{\Lambda _{\mathrm{2}},\Lambda }_{\mathrm{2}\mathcal{C}},\mathbf{%
\tilde{\Lambda}}_{\mathrm{2}\mathcal{C}},\mathbf{\Lambda }_{\mathrm{2}%
}^{\ast })$ can be drawn, specifically the following properties$:$

\begin{figure}[tbph]
\begin{center}
\includegraphics[width=14cm]{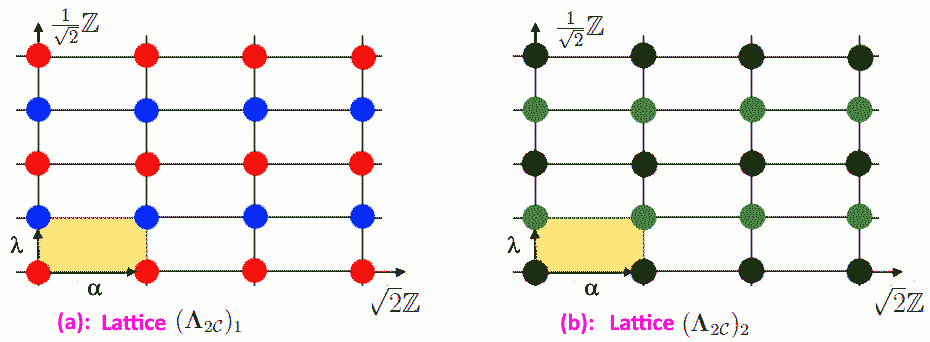}
\end{center}
\par
\vspace{-0.5cm}
\caption{The structure of the lattices $(\mathbf{\Lambda }_{\mathrm{2}%
\mathcal{C}})_{\mathrm{1}}$ and $(\mathbf{\Lambda }_{\mathrm{2}\mathcal{C}%
})_{\mathrm{2}}$ with sites shown in two different colors. The superposition
$(\mathbf{\Lambda }_{\mathrm{2}\mathcal{C}})_{\mathrm{1}}\cup (\mathbf{%
\Lambda }_{\mathrm{2}\mathcal{C}})_{\mathrm{2}}$ gives $\mathbf{\Lambda }_{%
\mathrm{2}}^{\ast }$. The even self dual lattice contains the $\mathbf{%
\Lambda }_{\mathrm{2}}$ made by red sites. Sites in blue belong to $(\mathbf{%
\Lambda }_{\mathrm{2}\mathcal{C}})_{\mathrm{1}}/\mathbf{\Lambda }_{\mathrm{2}%
}\simeq \mathbf{\Lambda }_{\mathrm{2}}+\mathbf{\protect\lambda }$. }
\label{3F}
\end{figure}

\textbf{A) the inclusion relations}\newline
First, notice that due to the homomorphism $SU(2)\times SU(2)^{\prime
}\simeq SO(4),$ the quartet $(\mathbf{\Lambda _{\mathrm{2}},\Lambda }_{%
\mathrm{2}\mathcal{C}},\mathbf{\tilde{\Lambda}}_{\mathrm{2}\mathcal{C}},%
\mathbf{\Lambda }_{\mathrm{2}}^{\ast })$ corresponds to lattice structures
associated with $SO(4).$ Moreover, using the embeddings $R^{\mathbf{su}%
_{2}}\subset W^{\mathbf{su}_{2}}$ and $R^{\prime \mathbf{su}_{2}}\subset
W^{\prime \mathbf{su}_{2}},$ it follows that:

\begin{itemize}
\item \emph{the inclusion relations}%
\begin{equation}
\mathbf{\Lambda }_{\mathrm{2}}\mathbf{\quad \subset \quad \Lambda }_{\mathrm{%
2}\mathcal{C}}\quad \subset \quad \mathbf{\Lambda }_{\mathrm{2}}^{\ast
}\qquad ,\qquad \mathbf{\Lambda }_{\mathrm{2}}\mathbf{\quad \subset \quad
\tilde{\Lambda}}_{\mathrm{2}\mathcal{C}}\quad \subset \quad \mathbf{\Lambda }%
_{\mathrm{2}}^{\ast }
\end{equation}%
admit different possible realisations for the even self dual lattice\ within
$\mathbf{\Lambda }_{\mathrm{2}}^{\ast }$. Since the weight lattices
decompose as $W^{\mathbf{su}_{2}}=\mathbb{A}\tbigcup \mathbb{B}$ with $%
\mathbb{A}\simeq \mathbb{B}\simeq R^{\mathbf{su}_{2}},$ and similarly for $%
W^{\prime \mathbf{su}_{2}}=\mathbb{A}^{\prime }\tbigcup \mathbb{B}^{\prime
}, $ the intermediate lattice $\mathbf{\Lambda }_{\mathrm{2}\mathcal{C}}$
can be realised either as $\mathbb{A}\times W^{\prime \mathbf{su}_{2}}$ or
like $\mathbb{B}\times W^{\prime \mathbf{su}_{2}}$. An analogous structure
takes place for $\mathbf{\tilde{\Lambda}}_{\mathrm{2}\mathcal{C}}$ which can
be similarly factorised either as $W^{\mathbf{su}_{2}}\times \mathbb{A}%
^{\prime }$ or like $W^{\mathbf{su}_{2}}\times \mathbb{B}^{\prime }$. Hence,
there are 2+2 configurations for the even self dual sublattices as follows%
\begin{equation}
\begin{tabular}{ccccc}
&  & ${\small [\mathbf{\Lambda }_{\mathrm{2}\mathcal{C}}]}_{1}$ &  &  \\
& $\nearrow $ &  & $\searrow $ &  \\
$\mathbf{\Lambda }_{\mathrm{2}}$ &  &  &  & $\mathbf{\Lambda }_{\mathrm{2}%
}^{\ast }$ \\
& $\searrow $ &  & $\nearrow $ &  \\
&  & ${\small [\mathbf{\Lambda }_{\mathrm{2}\mathcal{C}}]}_{2}$ &  &
\end{tabular}%
\qquad ,\qquad
\begin{tabular}{ccccc}
&  & ${\small [\mathbf{\tilde{\Lambda}}_{\mathrm{2}\mathcal{C}}]}_{1}$ &  &
\\
& $\nearrow $ &  & $\searrow $ &  \\
$\mathbf{\Lambda }_{\mathrm{2}}$ &  &  &  & $\mathbf{\Lambda }_{\mathrm{2}%
}^{\ast }$ \\
& $\searrow $ &  & $\nearrow $ &  \\
&  & ${\small [\mathbf{\tilde{\Lambda}}_{\mathrm{2}\mathcal{C}}]}_{2}$ &  &
\end{tabular}
\label{d1}
\end{equation}%
with
\begin{equation}
\begin{tabular}{lll}
${\small [\mathbf{\Lambda }_{\mathrm{2}\mathcal{C}}]}_{1}$ & $=$ & $\mathbb{A%
}\times W^{\prime \mathbf{su}_{2}}$ \\
${\small [\mathbf{\Lambda }_{\mathrm{2}\mathcal{C}}]}_{2}$ & $=$ & $\mathbb{B%
}\times W^{\prime \mathbf{su}_{2}}$%
\end{tabular}%
\qquad ,\qquad
\begin{tabular}{lll}
${\small [\mathbf{\tilde{\Lambda}}_{\mathrm{2}\mathcal{C}}]}_{1}$ & $=$ & $%
W^{\mathbf{su}_{2}}\times \mathbb{A}^{\prime }$ \\
${\small [\mathbf{\tilde{\Lambda}}_{\mathrm{2}\mathcal{C}}]}_{2}$ & $=$ & $%
W^{\mathbf{su}_{2}}\times \mathbb{B}^{\prime }$%
\end{tabular}%
\end{equation}%
Notice that the four ${\small [\mathbf{\Lambda }_{\mathrm{2}\mathcal{C}}]}%
_{i}$ and ${\small [\mathbf{\tilde{\Lambda}}_{\mathrm{2}\mathcal{C}}]}_{j}$
constitute 2-component multiplets of the discriminant $\mathbb{Z}_{2}\times
\mathbb{Z}_{2}^{\prime }.$

\item \emph{the remarkable superposition property}%
\begin{equation}
\mathbf{\Lambda }_{\mathrm{2}}^{\ast }={\small [\mathbf{\Lambda }_{\mathrm{2}%
\mathcal{C}}]}_{1}\text{ }\bigcup \text{ }{\small [\mathbf{\Lambda }_{%
\mathrm{2}\mathcal{C}}]}_{2}\qquad ,\qquad \widetilde{\mathbf{\Lambda }}_{%
\mathrm{2}}^{\ast }={\small [\mathbf{\tilde{\Lambda}}_{\mathrm{2}\mathcal{C}%
}]}_{1}\text{ }\bigcup \text{ }{\small [\mathbf{\tilde{\Lambda}}_{\mathrm{2}%
\mathcal{C}}]}_{2}
\end{equation}%
leads to 2+2 isomorphic self dual lattices ${\small [\mathbf{\Lambda }_{%
\mathrm{2}\mathcal{C}}]}_{i}$ and ${\small [\mathbf{\tilde{\Lambda}}_{%
\mathrm{2}\mathcal{C}}]}_{i}$ within the dual lattice $\mathbf{\Lambda }_{%
\mathrm{2}}^{\ast }.$ The even self dual lattices ${\small [\mathbf{\Lambda }%
_{\mathrm{2}\mathcal{C}}]}_{i}$ form a doublet of $\mathbb{Z}_{2}$
isomorphic to the discriminant $W_{\mathrm{2}}^{\ast }/R_{\mathrm{2}}$ while
the analogous even self duals ${\small [\mathbf{\tilde{\Lambda}}_{\mathrm{2}%
\mathcal{C}}]}_{i}$ form a doublet of $\mathbb{Z}_{2}^{\prime }\simeq W_{%
\mathrm{2}}^{\prime \ast }/R_{\mathrm{2}}^{\prime }.$ A graphical depiction
of $\mathbf{\Lambda }_{\mathrm{2}}^{\ast }$ is shown in Figure \textbf{\ref%
{f2}} where the unit cell spanned by $\mathbf{\lambda }\wedge \mathbf{%
\lambda }^{\prime }=\frac{1}{2}\mathbf{u}\wedge \mathbf{u}^{\prime }$ is
highlighted in yellow.
\begin{figure}[tbph]
\begin{center}
\includegraphics[width=10cm]{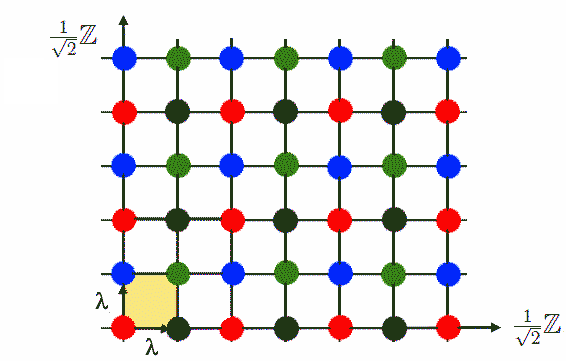}
\end{center}
\par
\vspace{-0.5cm}
\caption{Weight lattice $\mathbf{\Lambda }_{\mathrm{2}}^{\ast }:$ It is a 2D
lattice given by the cross product of two weight lattices of SU(2). Site
positions of these lattices have four different colors. The sites in each 1D
straightline in $\mathbf{\Lambda }_{\mathrm{2}}^{\ast }$ (including
horizontal and vertical axes ) have two colors.}
\label{f2}
\end{figure}
The four color scheme used for the weight lattice $\mathbf{\Lambda }_{%
\mathrm{2}}^{\ast }$ illustrates the coset structure $\mathbf{\Lambda }_{%
\mathrm{2}}^{\ast }/\mathbf{\Lambda _{\mathrm{2}}}\simeq $ $\mathbb{Z}%
_{2}\times \mathbb{Z}_{2}$ descending from $W^{\mathbf{su}_{2}}/R^{\mathbf{su%
}_{2}}\simeq \mathbb{Z}_{2}$ and (\ref{L}). The graphic representation of
the sublattices $\mathbf{\Lambda _{\mathrm{2}}}$, $\mathbf{\Lambda }_{%
\mathrm{2}\mathcal{C}}$ and $\mathbf{\tilde{\Lambda}}_{\mathrm{2}\mathcal{C}%
} $ will be presented later; in Figures \textbf{\ref{3F} }and\textbf{\ \ref%
{4F}}.
\end{itemize}

\

\textbf{B) }${\small [}\mathbf{\Lambda }{\small _{\mathrm{2}\mathcal{C}}]}_{%
\mathrm{j}}$ and $\mathbf{\Lambda }_{\mathrm{2}}^{\ast }$\textbf{\ as
superpositions of sublattices isomorphic to }$\mathbf{\Lambda }_{\mathrm{2}}$%
\newline
By applying the decomposition W$^{\mathbf{su}_{2}}=\mathbb{A}\cup \mathbb{B}$
to (\ref{L}), we find that both $\mathbf{\Lambda }_{\mathrm{2}}^{\ast }$ and
the lattices ${\small [\mathbf{\Lambda }_{\mathrm{2}\mathcal{C}}]}_{\mathrm{j%
}}$,${\small [\mathbf{\tilde{\Lambda}}_{\mathrm{2}\mathcal{C}}]}_{\mathrm{j}%
} $ can be written as superpositions of various 2D sublattices, all
isomorphic to $\mathbf{\Lambda }_{\mathrm{2}}$. Specifically, $\mathbf{%
\Lambda }_{\mathrm{2}}^{\ast }$ splits as%
\begin{eqnarray}
\mathbf{\Lambda }_{\mathrm{2}}^{\ast } &=&\left( \mathbb{A}\times \mathbb{A}%
^{\prime }\right) \ \ \ \bigcup \ \ \ \left( \mathbb{A}\times \mathbb{B}%
^{\prime }\right) \ \ \ \bigcup \ \ \ \left( \mathbb{B}\times \mathbb{A}%
^{\prime }\right) \ \ \ \bigcup \ \ \ \left( \mathbb{B}\times \mathbb{B}%
^{\prime }\right)  \notag \\
&\simeq &\ \left( \mathbf{\Lambda }_{\mathrm{2}}\right) _{\mathrm{11}}\ \ \
\ \ \bigcup \ \ \ \ \left( \mathbf{\Lambda }_{\mathrm{2}}\right) _{\mathrm{12%
}}\ \ \ \bigcup \ \ \ \ \ \left( \mathbf{\Lambda }_{\mathrm{2}}\right) _{%
\mathrm{21}}\ \ \ \ \bigcup \ \ \ \left( \mathbf{\Lambda }_{\mathrm{2}%
}\right) _{\mathrm{22}}  \label{et}
\end{eqnarray}%
with $\left( \mathbf{\Lambda }_{\mathrm{2}}\right) _{\mathrm{ij}}\simeq
\mathbb{A}_{\mathrm{i}}\times \mathbb{A}_{\mathrm{j}}$ where we set $\mathbb{%
A}_{\mathrm{1}}=\mathbb{A}$ ($\simeq \mathbb{A}^{\prime }$) and $\mathbb{A}_{%
\mathrm{2}}=\mathbb{B}$ ($\simeq \mathbb{B}^{\prime }$)$.$ The unit cell of $%
\mathbf{\Lambda }_{\mathrm{2}}^{\ast }$ has area $\mathbf{\lambda }\wedge
\mathbf{\lambda }^{\prime }=\frac{1}{2}\mathbf{u}\wedge \mathbf{u}^{\prime }$
normalised as $uc\mathbf{\Lambda }_{\mathrm{2}}^{\ast }=1/2.$ Due to the
isomorphisms $\mathbb{B}=\mathbb{A}+\mathbf{\lambda }$ and $\mathbb{B}%
^{\prime }=\mathbb{A}^{\prime }+\mathbf{\lambda }^{\prime }$, all four
sublattices $\left( \mathbf{\Lambda }_{\mathrm{2}}\right) _{\mathrm{ij}}$ in
(\ref{et}) are mutually isomorphic. In particular, amongst others, we find:%
\begin{equation}
\mathbb{A}\times \mathbb{A}^{\prime }\simeq \mathbb{B}\times \mathbb{B}%
^{\prime }\qquad ,\qquad \mathbb{A}\times \mathbb{B}^{\prime }\simeq \mathbb{%
B}\times \mathbb{A}^{\prime }
\end{equation}%
Moreover, the even self duals ${\small [\mathbf{\Lambda }_{\mathrm{2}%
\mathcal{C}}]}_{\mathrm{i}}$ and ${\small [\mathbf{\tilde{\Lambda}}_{\mathrm{%
2}\mathcal{C}}]}_{\mathrm{i}}$ are realised as follows
\begin{eqnarray}
{\small [\mathbf{\Lambda }_{\mathrm{2}\mathcal{C}}]}_{\mathrm{i}} &=&\text{
\ }\mathbb{A}_{\mathrm{i}}\text{ \ }\times \text{ \ }W^{\prime \mathbf{su}%
_{2}}  \notag \\
&=&\left( \mathbb{A}_{\mathrm{i}}\times \mathbb{A}^{\prime }\right) \text{ \
}\bigcup \text{ \ }\left( \mathbb{A}_{\mathrm{i}}\times \mathbb{B}^{\prime
}\right)  \label{E} \\
&&  \notag \\
{\small [\mathbf{\tilde{\Lambda}}_{\mathrm{2}\mathcal{C}}]}_{\mathrm{i}} &=&%
\text{ \ }W^{\mathbf{su}_{2}}\text{ \ }\times \text{ \ }\mathbb{A}_{\mathrm{i%
}}^{\prime }  \notag \\
&=&\left( \mathbb{A}\times \mathbb{A}_{\mathrm{i}}^{\prime }\right) \text{ \
}\bigcup \text{ \ }\left( \mathbb{B}\times \mathbb{A}_{\mathrm{i}}^{\prime
}\right)  \label{EE}
\end{eqnarray}%
These are isomorphic (${\small [\mathbf{\Lambda }_{\mathrm{2}\mathcal{C}}]}_{%
\mathrm{i}}\simeq {\small [\mathbf{\tilde{\Lambda}}_{\mathrm{2}\mathcal{C}}]}%
_{\mathrm{i}}$) with unit cells given by $\mathbf{\alpha }\wedge \mathbf{%
\lambda }^{\prime }=\mathbf{u}\wedge \mathbf{u}^{\prime }$ and normalised as
$uc(\mathbf{\Lambda }_{\mathrm{2}\mathcal{C}})_{\mathrm{i}}=uc(\mathbf{%
\tilde{\Lambda}}_{\mathrm{2}\mathcal{C}})_{\mathrm{i}}=1;$ that is twice the
fundamental area of the dual lattice: $uc(\mathbf{\Lambda }_{\mathrm{2}%
\mathcal{C}})_{\mathrm{i}}=2uc\mathbf{\Lambda }_{\mathrm{2}}^{\ast }.$ The
graphics of $(\mathbf{\Lambda }_{\mathrm{2}\mathcal{C}})_{\mathrm{i}}$ are
depicted in the \textbf{Figure} \textbf{\ref{3F} }with lattice sites colored
red and blue and corresponding sites for $(\mathbf{\tilde{\Lambda}}_{\mathrm{%
2}\mathcal{C}})_{\mathrm{i}}$\textrm{\ }are marked in black and green.
From the above splittings, we deduce the following: $\left( \mathbf{i}%
\right) $ $\mathbf{\Lambda }_{\mathrm{2}}^{\ast }$ consists of four 2D
sublattices\ given by $\mathbb{A}\times \mathbb{A}^{\prime }\ $and$\ \mathbb{%
A}\times \mathbb{B}^{\prime }\ $as well as $\mathbb{B}\times \mathbb{A}%
^{\prime }\ $and $\mathbb{B}\times \mathbb{B}^{\prime }$; while the even
self duals $(\mathbf{\Lambda }_{\mathrm{2}\mathcal{C}})_{\mathrm{i}}$ and
the $(\mathbf{\tilde{\Lambda}}_{\mathrm{2}\mathcal{C}})_{\mathrm{i}}$\ have
two superposed sublattices given by (\ref{E}-\ref{EE}). $\left( \mathbf{ii}%
\right) $ The dual $\mathbf{\Lambda }_{\mathrm{2}}^{\ast }$ is given by the
superposition $(\mathbf{\Lambda }_{\mathrm{2}\mathcal{C}})_{\mathrm{1}}\cup (%
\mathbf{\Lambda }_{\mathrm{2}\mathcal{C}})_{\mathrm{2}}.$

\begin{figure}[tbph]
\begin{center}
\includegraphics[width=16cm]{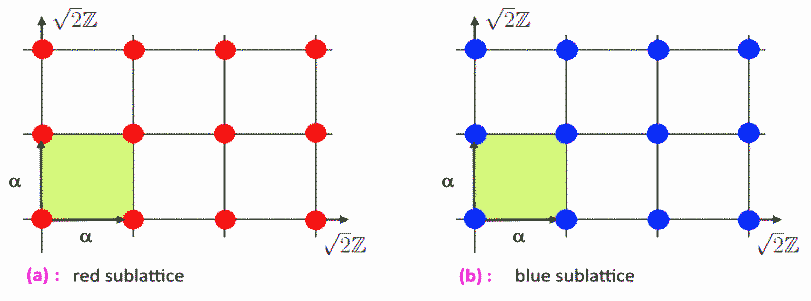}
\end{center}
\par
\vspace{-0.5cm}
\caption{Two isomorphic lattices $\mathbf{\Lambda }_{\mathrm{2}}$; (\textbf{a%
}) red sublattice of $\mathbf{\Lambda }_{\mathrm{2}}^{\ast }$. (\textbf{b})
blue sublattice of $\mathbf{\Lambda }_{\mathrm{2}}^{\ast }$. The two
additional ones are given by the black and the green sublattices of $\mathbf{%
\Lambda }_{\mathrm{2}}^{\ast }$ in the \textbf{Figures \protect\ref{f2}} and
\textbf{\protect\ref{3F}}. They are isomorphic to the root lattice of SO(4)
given by $R^{\mathbf{su}_{2}}\times R^{\prime \mathbf{su}_{2}}$. The area of
its normalised unit cell is equal to $uc\mathbf{\Lambda }_{\mathrm{2}}=2.$}
\label{4F}
\end{figure}

Regarding the even lattice $\mathbf{\Lambda }_{\mathrm{2}},$ it describes
the root lattice of $SO(4)$ and it appears as a 2D sublattice in each one of
$(\mathbf{\Lambda }_{\mathrm{2}\mathcal{C}})_{\mathrm{i}}$ and $(\mathbf{%
\tilde{\Lambda}}_{\mathrm{2}\mathcal{C}})_{\mathrm{i}}$ as well as of $%
\mathbf{\Lambda }_{\mathrm{2}}^{\ast }$ as schematically illustrated in (\ref%
{d1}). A visualisation of $\mathbf{\Lambda }_{\mathrm{2}}$ is depicted in
Figure \textbf{\ref{4F}.}
The unit cell in this lattice is defined by $\mathbf{\alpha }\wedge \mathbf{%
\alpha }^{\prime }=2\mathbf{u}\wedge \mathbf{u}^{\prime },$ giving the
normalised area:
\begin{equation}
uc\mathbf{\Lambda }_{\mathrm{2}}=2uc\mathbf{\Lambda }_{\mathrm{2}\mathcal{C}%
}=4uc\mathbf{\Lambda }_{\mathrm{2}}^{\ast }=2
\end{equation}

\subsection{Generalisation to higher dimensional $\mathbf{\Lambda _{\mathrm{k%
}}},\mathbf{\Lambda }_{\mathrm{k}\mathcal{C}},\mathbf{\Lambda }_{\mathrm{k}%
}^{\ast }$}

Here, we describe two natural extensions of the above 2D lattices $(\mathbf{%
\Lambda _{\mathrm{2}},}{\small [}\mathbf{\Lambda }{\small _{\mathrm{2}%
\mathcal{C}}]}_{\mathrm{j=1,2}},\mathbf{\Lambda }_{\mathrm{2}}^{\ast }).$
The first concerns the 2D lattices $\mathbf{\Lambda }_{\mathrm{k}},{\small [}%
\mathbf{\Lambda }{\small _{\mathrm{k}\mathcal{C}}]}_{\mathrm{j}},\mathbf{%
\Lambda }_{\mathrm{k}}^{\ast }$ indexed by the Chern-Simons level $\mathrm{k}%
>2$ with fixed dimension $\dim \mathbf{\Lambda }_{\mathrm{k}}=\dim \mathbf{%
\Lambda }{\small _{\mathrm{k}\mathcal{C}}}=\dim \mathbf{\Lambda }_{\mathrm{k}%
}^{\ast }=2$. These may be re-expressed as:
\begin{equation}
\mathbf{\Lambda }_{\mathrm{k}}=\mathbf{\Lambda }_{\mathrm{k}}^{(2,0)}\quad
,\qquad {\small [}\mathbf{\Lambda }{\small _{\mathrm{k}\mathcal{C}}]}_{%
\mathrm{j}}={\small [}\mathbf{\Lambda }_{\mathrm{k}\mathcal{C}}^{(1,1)}%
{\small ]}_{\mathrm{j}}\quad ,\qquad \mathbf{\Lambda }_{\mathrm{k}}^{\ast }=%
\mathbf{\Lambda }_{\mathrm{k}}^{(0,2)}
\end{equation}%
They obey the embedding $\mathbf{\Lambda }_{\mathrm{k}}^{(2,0)}\subset
{\small [}\mathbf{\Lambda }_{\mathrm{k}\mathcal{C}}^{(1,1)}{\small ]}_{%
\mathrm{j}}\subset \mathbf{\Lambda }_{\mathrm{k}}^{(0,2)}$ along with the
superposition property $\mathbf{\Lambda }_{\mathrm{k}}^{(0,2)}=\tbigcup%
\nolimits_{\mathrm{j}}{\small [}\mathbf{\Lambda }_{\mathrm{k}\mathcal{C}%
}^{(1,1)}{\small ]}_{\mathrm{j}}.$ \newline
The second generalisation involves higher dimensional lattices, constructed
via iterated tensor products:%
\begin{equation}
\mathbf{\Lambda }_{\mathrm{k}}=\mathbf{\Lambda }_{\mathrm{k}}^{(2d,0)}\quad
,\qquad {\small [}\mathbf{\Lambda }{\small _{\mathrm{k}\mathcal{C}}]}_{%
\mathrm{j}}={\small [}\mathbf{\Lambda }_{\mathrm{k}\mathcal{C}}^{(d,d)}%
{\small ]}_{\mathrm{\gamma }}\quad ,\qquad \mathbf{\Lambda }_{\mathrm{k}%
}^{\ast }=\mathbf{\Lambda }_{\mathrm{k}}^{(0,2d)}
\end{equation}%
with discriminant $\mathbf{\Lambda }_{\mathrm{k}}^{\ast }/\mathbf{\Lambda }_{%
\mathrm{k}}\simeq \mathbb{Z}_{\mathrm{k}}^{d}.$ They satisfy the inclusion
relations $\mathbf{\Lambda }_{\mathrm{k}}^{(2d,0)}\subset {\small [}\mathbf{%
\Lambda }_{\mathrm{k}\mathcal{C}}^{(d,d)}{\small ]}_{\mathrm{\gamma }%
}\subset \mathbf{\Lambda }_{\mathrm{k}}^{(0,2d)}$ in addition to $\mathbf{%
\Lambda }_{\mathrm{k}}^{(0,2d)}=\tbigcup\nolimits_{\mathrm{\gamma }}{\small [%
}\mathbf{\Lambda }_{\mathrm{k}\mathcal{C}}^{(d,d)}{\small ]}_{\mathrm{\gamma
}}.$

\subsubsection{Revisiting the 2D lattices $\mathbf{\Lambda _{\mathrm{k}}},%
\mathbf{\Lambda }_{\mathrm{k}\mathcal{C}},\mathbf{\Lambda }_{\mathrm{k}%
}^{\ast }$}

For levels \textrm{k\TEXTsymbol{>}2}, the previous one dimensional weight W$%
^{\mathbf{su}_{2}}$ and the root R$^{\mathbf{su}_{2}}$ lattices (associated
with \textrm{k=2 }and now denoted as\textrm{\ }W$_{\mathrm{2}}^{\mathbf{su}%
_{2}}$ and R$_{\mathrm{2}}^{\mathbf{su}_{2}}$) generalize to:%
\begin{equation}
W_{\mathrm{k}}^{\mathbf{su}_{2}}=\left\{ \mathbf{w}=\frac{m}{\sqrt{\mathrm{k}%
}}\mathbf{u};\quad m\in \mathbb{Z}\right\} \quad ,\quad R_{\mathrm{k}}^{%
\mathbf{su}_{2}}=\left\{ \mathbf{r}=\sqrt{\mathrm{k}}n\mathbf{u};\quad n\in
\mathbb{Z}\right\}
\end{equation}%
where $\mathbf{u}$ is a unit vector as before. They will be referred to as
weight-like and root-like lattices. From their expressions, we see that $%
\mathbf{r}=\mathrm{k}\times \mathbf{w}$ (i.e: $n=\mathrm{k}m$) consistent
with $W_{\mathrm{k}}^{\mathbf{su}_{2}}\ \subset R_{\mathrm{k}}^{\mathbf{su}%
_{2}}.$ This leads to the discriminant group:
\begin{equation}
W_{\mathrm{k}}^{\mathbf{su}_{2}}/R_{\mathrm{k}}^{\mathbf{su}_{2}}\simeq
\mathbb{Z}/\mathrm{k}\mathbb{Z\simeq Z}_{\mathrm{k}}
\end{equation}%
extending the standard $W_{\mathrm{2}}^{\mathbf{su}_{2}}/R_{\mathrm{2}}^{%
\mathbf{su}_{2}}\simeq \mathbb{Z}_{\mathrm{2}}$ of the su(2) lattices.
Moreover, the weight-like lattice $W_{\mathrm{k}}^{\mathbf{su}_{2}}$\
decomposes as a union of k isomorphic 1D sublattices $\mathbb{A}_{\mathrm{j}%
} $ as follows
\begin{equation}
W_{\mathrm{k}}^{\mathbf{su}_{2}}=\text{ }\mathbb{A}_{\mathrm{1}}\text{ }%
\bigcup \text{ }\mathbb{A}_{\mathrm{2}}\text{ }\bigcup \text{ }...\text{ }%
\bigcup \text{ }\mathbb{A}_{\mathrm{k}}
\end{equation}%
with%
\begin{equation}
\mathbb{A}_{\mathrm{j}}\simeq R_{\mathrm{k}}^{\mathbf{su}_{2}}\qquad ,\qquad
\mathrm{j}\in \left\{ 1,...,\mathrm{k}\right\}
\end{equation}%
The corresponding 2D lattices $\mathbf{\Lambda }_{\mathrm{k}},$ $\mathbf{%
\Lambda }_{\mathrm{k}}^{\ast },$ ${\small [}\mathbf{\Lambda }_{\mathrm{k}%
\mathcal{C}}{\small ]}_{\mathrm{j}}$ are respectively defined by the cross
product $\mathbf{\Lambda }_{\mathrm{k}}=R_{\mathrm{k}}^{\mathbf{su}%
_{2}}\times R_{\mathrm{k}}^{\prime \mathbf{su}_{2}}$ and%
\begin{equation}
\begin{tabular}{lll}
$\mathbf{\Lambda }_{\mathrm{k}}^{\ast }$ & $=$ & $W_{\mathrm{k}}^{\mathbf{su}%
_{2}}\times W_{\mathrm{k}}^{\prime \mathbf{su}_{2}}$ \\
$W_{\mathrm{k}}^{\mathbf{su}_{2}}$ & $=$ & $\tbigcup\limits_{\mathrm{j=1}}^{%
\mathrm{k}}\text{ }\mathbb{A}_{\mathrm{j}}$%
\end{tabular}%
\qquad ,\qquad
\begin{tabular}{lll}
${\small [}\mathbf{\Lambda }_{\mathrm{k}\mathcal{C}}{\small ]}_{\mathrm{j}}$
& $=$ & $\mathbb{A}_{\mathrm{j}}\times W_{\mathrm{k}}^{\prime \mathbf{su}%
_{2}}$ \\
$\mathbb{A}_{\mathrm{j}}$ & $\simeq $ & $R_{\mathrm{k}}^{\mathbf{su}_{2}}$%
\end{tabular}%
\end{equation}%
with
\begin{equation}
\begin{tabular}{|c|ccc|c|c|c|}
\hline
lattice & \multicolumn{3}{|c|}{site vectors} & $\left\langle \mathbf{\ast }.%
\mathbf{\ast }\right\rangle _{\text{\textsc{e}}}$ & $\left\langle \mathbf{%
\ast }.\mathbf{\ast }\right\rangle _{\text{\textsc{l}}}$ & {\small unit cell}
\\ \hline
$\mathbf{\Lambda }_{\mathrm{k}}^{\ast }$ & $\mathbf{w}$ & $=$ & $\frac{1}{%
\sqrt{\mathrm{k}}}\left( m^{1}\mathbf{u}_{1}+m^{2}\mathbf{u}_{2}\right) $ & $%
\frac{1}{\mathrm{k}}\left( m_{1}\right) ^{2}+\frac{1}{\mathrm{k}}\left(
m_{2}\right) ^{2}$ & $\frac{2}{\mathrm{k}}m_{2}m_{1}$ & $\frac{1}{\mathrm{k}}
$ \\
$\mathbf{\Lambda }_{\mathrm{k}}$ & $\mathbf{r}$ & $=$ & $\sqrt{\mathrm{k}}%
\left( n^{1}\mathbf{u}_{1}+n^{2}\mathbf{u}_{2}\right) $ & $\mathrm{k}\left(
n_{1}\right) ^{2}+\mathrm{k}\left( n_{2}\right) ^{2}$ & $2\mathrm{k}%
n_{2}n_{1}$ & $\mathrm{k}$ \\
${\small [}\mathbf{\Lambda }_{\mathrm{k}\mathcal{C}}{\small ]}_{\mathrm{j}}$
& $\mathbf{x}$ & $=$ & $\sqrt{\mathrm{k}}n^{1}\mathbf{u}_{1}+\frac{1}{\sqrt{%
\mathrm{k}}}m^{2}\mathbf{u}_{2}$ & $\mathrm{k}\left( n_{1}\right) ^{2}+\frac{%
1}{\mathrm{k}}\left( m_{2}\right) ^{2}$ & $2n_{2}m_{1}$ & $1$ \\ \hline
\end{tabular}
\label{2k}
\end{equation}%
\begin{equation*}
\end{equation*}%
As for their normalised unit cells, these 2D lattices depend on the level k
and obey the inclusions $\mathbf{\Lambda }_{\mathrm{k}}\subset {\small [}%
\mathbf{\Lambda }_{\mathrm{k}\mathcal{C}}{\small ]}_{\mathrm{j}}\subset
\mathbf{\Lambda }_{\mathrm{k}}^{\ast }$ with the isomorphisms ${\small [}%
\mathbf{\Lambda }_{\mathrm{k}\mathcal{C}}{\small ]}_{\mathrm{j}}\simeq
\mathbf{\Lambda }_{\mathrm{k}\mathcal{C}}$ as exhibited by the Figure
\textbf{\ref{3F}}. Graphically, we have%
\begin{equation}
\begin{tabular}{ccccc}
&  & $\left( \mathbf{\Lambda }_{\mathrm{k}\mathcal{C}}\right) _{\mathrm{1}}$
&  &  \\
& $\nearrow $ & $\vdots $ & $\searrow $ &  \\
$\mathbf{\Lambda }_{\mathrm{k}}$ & $\longrightarrow $ & $\vdots $ & $%
\longrightarrow $ & $\mathbf{\Lambda }_{\mathrm{k}}^{\ast }$ \\
& $\searrow $ & $\vdots $ & $\nearrow $ &  \\
&  & $\left( \mathbf{\Lambda }_{\mathrm{k}\mathcal{C}}\right) _{\mathrm{k}}$
&  &
\end{tabular}%
\qquad ,\qquad \left( \mathbf{\Lambda }_{\mathrm{k}\mathcal{C}}\right) _{%
\mathrm{1}}\simeq \cdots \simeq \left( \mathbf{\Lambda }_{\mathrm{k}\mathcal{%
C}}\right) _{\mathrm{k}}
\end{equation}%
Observe that for the trivial level k=1, we recover $W_{\mathrm{1}}^{\mathbf{%
su}_{2}}\simeq R_{1}^{\mathbf{su}_{2}}\simeq \mathbb{Z}.$ As such, all three
associated lattices coincide $\mathbf{\Lambda }_{\mathrm{1}}\simeq \mathbf{%
\Lambda }_{\mathrm{1}\mathcal{C}}\simeq \mathbf{\Lambda }_{\mathrm{1}}^{\ast
}\simeq \mathbb{Z\times Z}.$ Observe also that from the table (\ref{2k}),\
the Lorentzian pairings $\left\langle \mathbf{\ast }.\mathbf{\ast }%
\right\rangle _{\text{\textsc{l}}}$ have the following properties:%
\begin{equation}
\begin{tabular}{|c|c|c|c|c|}
\hline
Lattice & $\ \mathbf{\Lambda }_{\mathrm{k}}$ \  & $\ \mathbf{\Lambda }_{%
\mathrm{k}\mathcal{C}}$ $\ $ & $\ \mathbf{\tilde{\Lambda}}_{\mathrm{k}%
\mathcal{C}}$ $\ $ & $\ \mathbf{\Lambda }_{\mathrm{k}}^{\ast }$ $\ $ \\
\hline
pairing $\left\langle \mathbf{\ast }.\mathbf{\ast }\right\rangle _{\text{%
\textsc{l}}}$ & $\ 2\mathrm{k}\mathbb{Z}$ $\ $ & $\ 2\mathbb{Z}$ $\ $ & $\ 2%
\mathbb{Z}$ $\ $ & $\ \frac{2}{\mathrm{k}}\mathbb{Z}$ $\ $ \\ \hline
unit cell & $\mathrm{k}\geq 1$ & $1$ & $1$ & $\frac{1}{\mathrm{k}}\leq 1$ \\
\hline
\end{tabular}%
\end{equation}%
with $2\mathrm{k}\mathbb{Z}\subset 2\mathbb{Z}\subset \frac{2}{\mathrm{k}}%
\mathbb{Z}$ coinciding for the trivial value $\mathrm{k}=1;$ but\ differ for
$\mathrm{k}\geq 2.$ Finally, inserting the decomposition $W_{\mathrm{k}}^{%
\mathbf{su}_{2}}=$ $\bigcup_{\mathrm{i}=1}^{\mathrm{k}}$ $\mathbb{A}_{%
\mathrm{i}}$ into $\mathbf{\Lambda }_{\mathrm{k}}^{\ast }=W_{\mathrm{k}}^{%
\mathbf{su}_{2}}\times W_{\mathrm{k}}^{\prime \mathbf{su}_{2}}$, we find
that there are k possible self dual sublattices $\left( \mathbf{\Lambda }_{%
\mathrm{k}\mathcal{C}}\right) _{\mathrm{i}}$ each of the form:%
\begin{equation}
\left( \mathbf{\Lambda }_{\mathrm{k}\mathcal{C}}\right) _{\mathrm{i}}=%
\mathbb{A}_{\mathrm{i}}\times \bigcup_{\mathrm{j}=1}^{\mathrm{k}}\mathbb{A}_{%
\mathrm{j}}\qquad ,\qquad \mathrm{i=1,\cdots ,k}
\end{equation}%
with the isomorphisms $\mathbb{A}_{\mathrm{1}}\simeq \cdots \simeq \mathbb{A}%
_{\mathrm{k}}\simeq R_{\mathrm{k}}^{\mathbf{su}_{2}}.$

\subsubsection{Tensor products}

The extension of $\mathbf{\Lambda _{\mathrm{k}}},\mathbf{\Lambda }_{\mathrm{k%
}\mathcal{C}},\mathbf{\tilde{\Lambda}}_{\mathrm{k}\mathcal{C}},\mathbf{%
\Lambda }_{\mathrm{k}}^{\ast }$ as defined in eq(\ref{2k}) to higher
dimensional lattices (particularly for $\dim \mathbf{\Lambda }_{\mathrm{k}%
}^{\ast }=\dim \mathbf{\Lambda }_{\mathrm{k}\mathcal{C}}=\dim \mathbf{%
\Lambda }_{\mathrm{k}}=2\mathrm{d}$ with positive integer $\mathrm{d}$) can
be carried out in several ways.\ In what follows, we focus on two such
extensions labeled by the su(2) and su(3) algebras. Other possibilities
exist and can be classified according to finite dimensional Lie algebras g;
we will briefly comment on these in the discussion section. In this
subsection, we present a higher dimensional extension inspired from
construction A of code CFTs using the language of tensor products of su(2)
lattices. A second generalisation based on su(3) lattices and their tensor
products will be detailed in the next section.\newline
The $2\mathrm{d}$- $\dim $ version of eq(\ref{2k}) is given by the tensorial
constructions $\left( \mathbf{\Lambda }_{\mathrm{k}}^{\ast }\right)
^{\otimes \mathrm{d}},$ $\left( \mathbf{\Lambda }_{\mathrm{k}}\right)
^{\otimes \mathrm{d}}$ and $\left( \mathbf{\Lambda }_{\mathrm{k}}\right)
^{\otimes \mathrm{d}}$ with $\dim \mathbf{\Lambda }_{\mathrm{k}}=2\mathrm{d}%
; $ these are defined as summarised in the following table:%
\begin{equation}
\begin{tabular}{|c|ccc|c|c|c|}
\hline
lattice & \multicolumn{3}{|c|}{site vectors} & $\left\langle \mathbf{\ast },%
\mathbf{\ast }\right\rangle _{\text{\textsc{e}}}$ & $\left\langle \mathbf{%
\ast }.\mathbf{\ast }\right\rangle _{\text{\textsc{l}}}$ & {\small unit cell}
\\ \hline
$\left( \mathbf{\Lambda }_{\mathrm{k}}^{\ast }\right) ^{\mathrm{d}}$ & $%
\mathbf{w}$ & $=$ & $\sum\limits_{l=1}^{\mathrm{d}}\mathbf{w}_{l}$ & $%
\sum\limits_{l=1}^{\mathrm{d}}\left\langle \mathbf{w}_{l},\mathbf{w}%
_{l}\right\rangle _{\text{\textsc{e}}}$ & $\sum\limits_{l=1}^{\mathrm{d}%
}\left\langle \mathbf{w}_{l},\mathbf{w}_{l}\right\rangle _{\text{\textsc{l}}%
} $ & $\frac{1}{\mathrm{k}^{\mathrm{d}}}$ \\
$\left( \mathbf{\Lambda }_{\mathrm{k}}\right) ^{\mathrm{d}}$ & $\mathbf{r}$
& $=$ & $\sum\limits_{l=1}^{\mathrm{d}}\mathbf{r}_{l}$ & $\sum\limits_{l=1}^{%
\mathrm{d}}\left\langle \mathbf{r}_{l},\mathbf{r}_{l}\right\rangle _{\text{%
\textsc{e}}}$ & $\sum\limits_{l=1}^{\mathrm{d}}\left\langle \mathbf{r}_{l},%
\mathbf{r}_{l}\right\rangle _{\text{\textsc{l}}}$ & $\mathrm{k}^{\mathrm{d}}$
\\
$\left( \mathbf{\Lambda }_{\mathrm{k}\mathcal{C}}\right) ^{\mathrm{d}}$ & $%
\mathbf{x}$ & $=$ & $\sum\limits_{l=1}^{\mathrm{d}}\mathbf{x}_{l}$ & $%
\sum\limits_{l=1}^{\mathrm{d}}\left\langle \mathbf{x}_{l},\mathbf{x}%
_{l}\right\rangle _{\text{\textsc{e}}}$ & $\sum\limits_{l=1}^{\mathrm{d}%
}\left\langle \mathbf{x}_{l},\mathbf{x}_{l}\right\rangle _{\text{\textsc{l}}%
} $ & $1$ \\ \hline
\end{tabular}
\label{tab1}
\end{equation}%
\begin{equation*}
\end{equation*}%
with $\mathrm{k}^{\mathrm{d}}\geq 1.$ Explicitly, we have:
\begin{equation*}
\begin{tabular}{|c|c|c|c|}
\hline
{\small lattice} & {\small site vectors} & $\left\langle \mathbf{\ast }.%
\mathbf{\ast }\right\rangle _{\text{\textsc{e}}}$ & $\left\langle \mathbf{%
\ast }.\mathbf{\ast }\right\rangle _{\text{\textsc{l}}}$ \\ \hline
$\left( \mathbf{\Lambda }_{\mathrm{k}}^{\ast }\right) ^{\mathrm{d}}$ & $%
\frac{1}{\sqrt{\mathrm{k}}}\left( m^{2l-1}\mathbf{u}_{2l-1}+m^{2l}\mathbf{u}%
_{2l}\right) $ & $\frac{1}{\mathrm{k}}\sum\limits_{l=1}^{\mathrm{d}}\left[
\left( m_{2l-1}\right) ^{2}+\left( m_{2l}\right) ^{2}\right] $ & $\frac{2}{%
\mathrm{k}}\sum\limits_{l=1}^{\mathrm{d}}m_{2l}m_{2l-1}$ \\
$\left( \mathbf{\Lambda }_{\mathrm{k}}\right) ^{\mathrm{d}}$ & $\sqrt{%
\mathrm{k}}\left( n^{2l-1}\mathbf{u}_{2l-1}+n^{2l}\mathbf{u}_{2l}\right) $ &
$\mathrm{k}\sum\limits_{l=1}^{\mathrm{d}}\left[ \left( n_{2l-1}\right)
^{2}+\left( n_{2l}\right) ^{2}\right] $ & $2\mathrm{k}\sum\limits_{l=1}^{%
\mathrm{d}}n_{2l}n_{2l-1}$ \\
$\left( \mathbf{\Lambda }_{\mathrm{k}\mathcal{C}}\right) ^{\mathrm{d}}$ & $%
\sqrt{\mathrm{k}}n^{2l-1}\mathbf{u}_{2l-1}+\frac{1}{\sqrt{\mathrm{k}}}m^{2l}%
\mathbf{u}_{2l}$ & $\sum\limits_{l=1}^{\mathrm{d}}\left[ \mathrm{k}\left(
n_{2l-1}\right) ^{2}+\frac{1}{\mathrm{k}}\left( m_{2l}\right) ^{2}\right] $
& $2\sum\limits_{l=1}^{\mathrm{d}}n_{2l}m_{2l-1}$ \\ \hline
\end{tabular}%
\end{equation*}%
\begin{equation*}
\end{equation*}%
By thinking about $\mathbf{\Lambda }_{\mathrm{k}\mathcal{C}}$ in terms of
the representations $\left( \mathbf{\Lambda }_{\mathrm{k}\mathcal{C}}\right)
_{\mathrm{j}}$ given by the fibration $\mathbb{A}_{\mathrm{j}}\times W_{%
\mathrm{k}},$ it results that $\left( \mathbf{\Lambda }_{\mathrm{k}\mathcal{C%
}}\right) ^{\otimes \mathrm{d}}=(\mathbb{A}_{\mathrm{j}})^{\otimes \mathrm{d}%
}\times \left( W_{\mathrm{k}}\right) ^{\otimes \mathrm{d}}.$

\section{SU(3)-Based Lattice Constructions}

In this section, we exploit structural properties of the 2D lattices of
su(3) Lie algebra to build 4D realisations of the inclusion relations $%
\mathbf{\Lambda }_{\mathrm{k}}^{\mathbf{su}_{3}}\subset \mathbf{\Lambda }_{%
\mathrm{k}\mathcal{C}}^{\mathbf{su}_{3}}\subset \mathbf{\Lambda }_{\mathrm{k}%
}^{\ast \mathbf{su}_{3}}$ for arbitrary integer Chern-Simons level integer $%
\mathrm{k}\geq 1:$
\begin{equation}
\dim \mathbf{\Lambda }_{\mathrm{k}}^{\mathbf{su}_{3}}=\dim \mathbf{\Lambda }%
_{\mathrm{k}\mathcal{C}}^{\mathbf{su}_{3}}=\dim \mathbf{\Lambda }_{\mathrm{k}%
}^{\ast \mathbf{su}_{3}}=4
\end{equation}%
The generalisation towards higher dimensions, namely $\dim \mathbf{\Lambda }%
_{\mathrm{k}}^{\mathbf{su}_{3}}=\dim \mathbf{\Lambda }_{\mathrm{k}\mathcal{C}%
}^{\mathbf{su}_{3}}=\dim \mathbf{\Lambda }_{\mathrm{k}}^{\ast \mathbf{su}%
_{3}}=4\mathrm{d}$, for positive integer \textrm{d}$\geq 2$ is
straightforward and follows by applying tensor product techniques, analogous
to those used in the table (\ref{tab1}) of the su(2) theory; it will be
briefly commented subsequently.

\subsection{Triangular/hexagonal lattices and extensions}

\begin{figure}[tbph]
\begin{center}
\includegraphics[width=16cm]{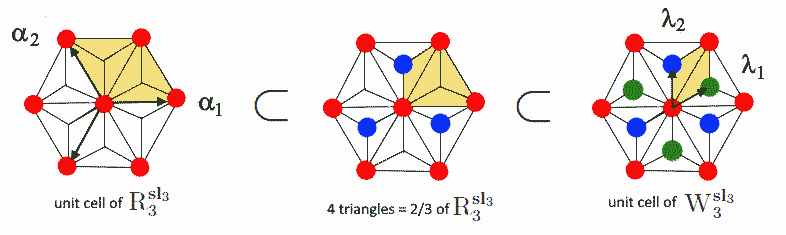}
\end{center}
\par
\vspace{-0.5cm}
\caption{On the left, a colored parallelogram (unit cell) of the hexagonal
lattice R$_{\mathrm{3}}^{\mathbf{su}_{3}}$. The 3 parallelograms (1 colored
and 2 uncolored) form altogether an hexagon with centre and vertices in red.
On the extreme right, we give the unit cell of W$_{\mathrm{3}}^{\mathbf{su}%
_{3}}$ with 3 colors$.$}
\label{2i}
\end{figure}

\begin{figure}[tbph]
\begin{center}
\includegraphics[width=12cm]{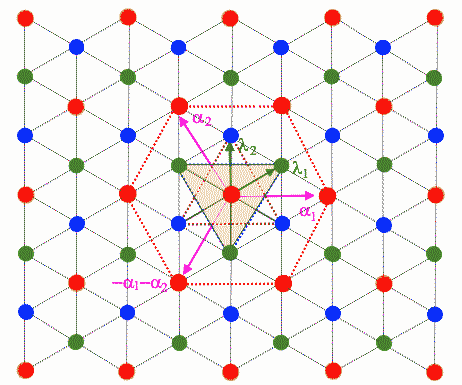}
\end{center}
\par
\vspace{-0.5cm}
\caption{Triangular lattice $\mathbb{Z}\mathbf{\protect\lambda }_{1}\oplus
\mathbb{Z}\mathbf{\protect\lambda }_{2}$ generated by the\textrm{\ SU(3)}
weights\textrm{\ }with $(\widehat{\mathbf{\protect\lambda }_{1},\mathbf{%
\protect\lambda }_{2}})=\protect\pi /3.$ It is the weight lattice \textsc{W}$%
_{\mathrm{3}}^{\mathbf{su}_{3}}$ of su(3). The unit cell is a triangle with
pistachio color. Sites in \textsc{W}$_{\mathrm{3}}^{\mathbf{su}_{3}}$ are
painted in three colors: red, blue and green.}
\label{T}
\end{figure}

The 2D triangular lattice and its hexagonal sublattice considered here
below\ are naturally interpreted in terms of the weight $W_{\mathrm{3}}^{%
\mathbf{su}_{3}}$ and roots $R_{\mathrm{3}}^{\mathbf{su}_{3}}$ lattices of
SU(3). They have the discriminant $W_{\mathrm{3}}^{\mathbf{su}_{3}}/R_{%
\mathrm{3}}^{\mathbf{su}_{3}}\simeq \mathbb{Z}_{3}$ \cite{R1,R2, R3}
signaling that the level of the associated CS theory is:
\begin{equation}
\mathrm{k=3}  \label{k3}
\end{equation}%
Site positions $\mathbf{w}_{\mathbf{m}}$ coordinating the weight lattice $W_{%
\mathrm{3}}^{\mathbf{su}_{3}}$ are labeled by integer 2-vectors $\mathbf{m}%
=(m_{1},m_{2});$ they are generated by the two fundamental weight vectors $%
\left\{ \mathbf{\lambda }^{i}\right\} _{i=1,2}$ of su(3) with intersection
matrix $K^{ij}=\mathbf{\lambda }^{i}.\mathbf{\lambda }^{j}.$ Explicitly, we
have
\begin{equation}
\mathbf{w}_{\mathbf{m}}=m_{1}\mathbf{\lambda }^{1}+m_{2}\mathbf{\lambda }%
^{2}\qquad ,\qquad (\widehat{\mathbf{\lambda }_{1},\mathbf{\lambda }_{2}})=%
\frac{2\pi }{6}
\end{equation}%
with unit cell $\mathbf{\lambda }_{1}\wedge \mathbf{\lambda }_{2}$ defining
an area of $ucW_{\mathrm{3}}^{\mathbf{su}_{3}}=\frac{1}{\sqrt{3}}.$ The root
$R_{\mathrm{3}}^{\mathbf{su}_{3}}\subset W_{\mathrm{3}}^{\mathbf{su}_{3}}$
forms a hexagonal sublattice with site vectors $\mathbf{r}_{\mathbf{n}}=n^{1}%
\mathbf{\alpha }_{1}+n^{2}\mathbf{\alpha }_{2}$ generated by the two simple
roots $\left\{ \mathbf{\alpha }_{1},\mathbf{\alpha }_{2}\right\} $ and inner
product\textrm{\ }$(\widehat{\mathbf{\alpha }_{1},\mathbf{\alpha }_{2}})=%
\frac{2\pi }{3}.$ For this 2D lattice $R_{\mathrm{3}}^{\mathbf{su}_{3}},$
the unit cell $\mathbf{\alpha }_{1}\wedge \mathbf{\alpha }_{2}$ has the area
$ucR_{\mathrm{3}}^{\mathbf{su}_{3}}=\sqrt{3}$ satisfying\textrm{\ }the
property%
\begin{equation}
\frac{ucR_{\mathrm{3}}^{\mathbf{su}_{3}}}{ucW_{\mathrm{3}}^{\mathbf{su}_{3}}}%
=3\qquad \Leftrightarrow \qquad \frac{\left\vert \mathbf{\alpha }_{1}\wedge
\mathbf{\alpha }_{2}\right\vert }{\left\vert \mathbf{\lambda }_{1}\wedge
\mathbf{\lambda }_{2}\right\vert }=3
\end{equation}%
This structure is illustrated in Figure \textbf{\ref{2i} }where a $R_{%
\mathrm{3}}^{\mathbf{su}_{3}\prime }$ hexagon (centre and vertices\ marked
in red) contains three fundamental unit cells (three parallelograms).
Compared to the weight lattice $W_{\mathrm{3}}^{\mathbf{su}_{3}}$, each unit
cell of $R_{\mathrm{3}}^{\mathbf{su}_{3}}$ comprises three unit cells of $W_{%
\mathrm{3}}^{\mathbf{su}_{3}}.$%
The intersection matrix $K_{ij}=\mathbf{\alpha }_{i}.\mathbf{\alpha }_{j}$
correspond to the SU(3) Cartan matrix, which is the inverse of $K^{ij};$
that is: $K_{il}K^{lj}=\delta _{i}^{j}.$ Explicitly, we have%
\begin{equation}
K_{ij}=\left(
\begin{array}{cc}
2 & -1 \\
-1 & 2%
\end{array}%
\right) \qquad ,\qquad K^{ij}=\left(
\begin{array}{cc}
\frac{2}{3} & \frac{1}{3} \\
\frac{1}{3} & \frac{2}{3}%
\end{array}%
\right)
\end{equation}%
with $\det K_{ij}=3$ and $\det K^{ij}=1/3.$ By using the relations $\mathbf{%
\lambda }_{1}=\frac{1}{3}\left( 2\mathbf{\alpha }_{1}+\mathbf{\alpha }%
_{2}\right) $ and $\mathbf{\lambda }_{2}=\frac{1}{3}\left( \mathbf{\alpha }%
_{1}+2\mathbf{\alpha }_{2}\right) ,$ we can also express weight vectors as
\begin{equation}
\mathbf{w}_{\mathbf{m}}=\frac{1}{3}\left( 2m_{1}+m_{2}\right) \mathbf{\alpha
}_{1}+\frac{1}{3}\left( m_{1}+2m_{2}\right) \mathbf{\alpha }_{2}
\end{equation}%
showing that $W_{\mathrm{3}}^{\mathbf{su}_{3}}$ contains indeed the root $R_{%
\mathrm{3}}^{\mathbf{su}_{3}}$\ as a sublattice (hexagonal) namely when $%
2m_{1}+m_{2}=3p$ and $m_{1}+2m_{2}=3q$ solved by $m_{1}=2p-q$ and $%
m_{2}=2q-p.$ Moreover, using the discriminant $W_{\mathrm{3}}^{\mathbf{su}%
_{3}}/R_{\mathrm{3}}^{\mathbf{su}_{3}}\simeq \mathbb{Z}_{3},$ we deduce that
$W_{\mathrm{3}}^{\mathbf{su}_{3}}$ is decomposable into a superposition of
three isomorphic 2D sublattices as given below,
\begin{equation}
W_{\mathrm{k}}^{\mathbf{su}_{3}}=\mathbb{A}\text{ }\bigcup \text{ }\mathbb{B}%
\text{ }\bigcup \text{ }\mathbb{C}\qquad with\qquad \mathrm{k=3}  \label{3}
\end{equation}%
with the isomorphisms property $\mathbb{A}\simeq \mathbb{B}\simeq \mathbb{C}%
\simeq R_{\mathrm{3}}^{\mathbf{su}_{3}}.$ A graphic representation of the
structure of\textrm{\ }\textsc{W}$_{\mathrm{3}}^{\mathbf{su}_{3}}$ is
depicted\textrm{\ }in Figure \textbf{\ref{T} }where the unit cell is
represented by the pistachio colored triangle\textrm{.}

\subsubsection{Models with generic CS level k}

These lattices, all of which are embedded in the weight lattice\textrm{\ }$%
W_{\mathrm{k}}^{\mathbf{su}_{3}}\simeq \mathbb{Z}\mathbf{\tilde{\lambda}}%
_{1}+\mathbb{Z}\mathbf{\tilde{\lambda}}_{2},$ can be constructed by
generalising the standard $W_{\mathrm{3}}^{\mathbf{su}_{3}}/R_{\mathrm{3}}^{%
\mathbf{su}_{3}}\simeq \mathbb{Z}_{3}$ and reinterpret the quotient as part
of a border family of discriminant groups defined as%
\begin{equation}
W_{\mathrm{k}}^{\mathbf{su}_{3}}/R_{\mathrm{k}}^{\mathbf{su}_{3}}\simeq
\mathbb{Z}_{\mathrm{k}}\qquad ,\qquad \mathrm{k}\in \mathbb{N}^{\ast }
\end{equation}%
with positive integer $\mathrm{k}$. Using the area of the unit cells $ucW_{%
\mathrm{k}}^{\mathbf{su}_{3}}=\frac{\sqrt{3}}{\mathrm{k}}$ and $ucR_{\mathrm{%
k}}^{\mathbf{su}_{3}}=\frac{\mathrm{k}}{\sqrt{3}},$ we find%
\begin{equation}
\frac{ucR_{\mathrm{k}}^{\mathbf{su}_{3}}}{ucW_{\mathrm{k}}^{\mathbf{su}_{3}}}%
=\frac{1}{3}\mathrm{k}^{2}>1  \label{kkk}
\end{equation}%
which holds for all integers $\mathrm{k}\geq 2.$ For this extension, the
site vectors $\mathbf{w}_{\mathbf{m}}$ of $W_{\mathrm{k}}^{\mathbf{su}_{3}}$
and $\mathbf{r}_{\mathbf{n}}$ of $R_{\mathrm{k}}^{\mathbf{su}_{3}}$ are
k-dependent; they generalise\textrm{\ }the case\textrm{\ }k=3 as follows:
\begin{equation}
\mathbf{w}_{\mathbf{m}}=\sqrt{\frac{3}{\mathrm{k}}}\left( m_{1}\mathbf{%
\lambda }^{1}+m_{2}\mathbf{\lambda }^{2}\right) \qquad \Leftrightarrow
\qquad \mathbf{w}_{\mathbf{m}}=m_{1}\mathbf{\tilde{\lambda}}^{1}+m_{2}%
\mathbf{\tilde{\lambda}}^{2}
\end{equation}%
in addition to%
\begin{equation}
\mathbf{r}_{\mathbf{n}}=\sqrt{\frac{\mathrm{k}}{3}}\left( n^{1}\mathbf{%
\alpha }_{1}+n^{2}\mathbf{\alpha }_{2}\right) \qquad \Leftrightarrow \qquad
\mathbf{r}_{\mathbf{n}}=n^{1}\mathbf{\tilde{\alpha}}_{1}+n^{2}\mathbf{\tilde{%
\alpha}}_{2}
\end{equation}%
with $\mathbf{\tilde{\alpha}}_{i}.\mathbf{\tilde{\alpha}}_{j}=\frac{\mathrm{k%
}}{3}K_{ij}$ and $\mathbf{\tilde{\lambda}}^{i}.\mathbf{\tilde{\lambda}}^{j}=%
\frac{3}{\mathrm{k}}K^{ij}$ being the 2$\times $2 matrices $K_{ij}$\ and $%
K^{ij}$\ as before. For generic integer values of $\mathrm{k}\geq 2,$ the $%
W_{\mathrm{k}}^{\mathbf{su}_{3}}$ is given by a superposition of k
sublattices $\mathbb{A}_{\mathrm{j}}$ as follows,
\begin{equation}
W_{\mathrm{k}}^{\mathbf{su}_{3}}=\mathbb{A}_{1}\text{ }\bigcup \text{ }%
\mathbb{A}_{2}\text{ }\cdots \bigcup \text{ }\mathbb{A}_{\mathrm{k}}
\label{k}
\end{equation}%
As for the\textrm{\ }$\mathbb{A},\mathbb{B}$ and $\mathbb{C}$ of eq(\ref{3})
for k=3, each $\mathbb{A}_{\mathrm{j}}$ is isomorphic to $R_{\mathrm{k}}^{%
\mathbf{su}_{3}}\simeq \mathbb{Z}\mathbf{\tilde{\alpha}}_{1}+\mathbb{Z}%
\mathbf{\tilde{\alpha}}_{2}.$ This description reveals several interesting
structural features which we discuss below:

\subsubsection{Three intervals for the level k}

Depending on the value of the Chern-Simons level k with respect to the
critical (\ref{k3}), we distinguish three regions:

\begin{description}
\item[$(\mathbf{i})$] \textbf{case }$\mathrm{k}=3$: It corresponds precisely
to the standard weight $W_{\mathrm{3}}^{\mathbf{su}_{3}}$ and root $R_{%
\mathrm{3}}^{\mathbf{su}_{3}}$ lattices of su(3). For this level, the
discriminant is given by $W_{\mathrm{3}}^{\mathbf{su}_{3}}/R_{\mathrm{3}}^{%
\mathbf{su}_{3}}\simeq \mathbb{Z}_{3}$ indicating that $W_{\mathrm{3}}^{%
\mathbf{su}_{3}}$ is given by the superposition $W_{\mathrm{3}}^{\mathbf{su}%
_{3}}=\mathbb{A}$ $\tbigcup $ $\mathbb{B}$ $\tbigcup $ $\mathbb{C}$
represented by the Figure \textbf{\ref{T} }with colored site positions: red
for $\mathbb{A}$, blue for $\mathbb{B}$ and green for $\mathbb{C}.$ The
three lattices are related by lattice translations such as:%
\begin{equation}
\begin{tabular}{lllllll}
$\mathbb{B}$ & $=$ & $\mathbb{A}-\mathbf{\lambda }_{1}$ & $\simeq $ & $%
\mathbb{A}+2\mathbf{\lambda }_{1}$ & $\simeq $ & $\mathbb{A}+\mathbf{\lambda
}_{2}$ \\
$\mathbb{C}$ & $=$ & $\mathbb{A}+\mathbf{\lambda }_{1}$ & $\simeq $ & $%
\mathbb{A}+2\mathbf{\lambda }_{2}$ & $\simeq $ & $\mathbb{A}-\mathbf{\lambda
}_{2}$%
\end{tabular}%
\end{equation}%
These equivalences are due to the relations $2\mathbf{\lambda }_{1}=\mathbf{%
\alpha }_{1}+\mathbf{\lambda }_{2}$ and $2\mathbf{\lambda }_{2}=\mathbf{%
\alpha }_{2}+\mathbf{\lambda }_{1}$ along with $\mathbf{\lambda }_{2}\simeq 2%
\mathbf{\lambda }_{1}\simeq -\mathbf{\lambda }_{1}$ and $\mathbf{\lambda }%
_{1}\simeq 2\mathbf{\lambda }_{2}\simeq -\mathbf{\lambda }_{2}$ seeing that $%
\mathbf{\lambda }_{1}+\mathbf{\lambda }_{2}=\mathbf{\alpha }_{1}+\mathbf{%
\alpha }_{2}$ belongs to the zero class $\bar{0}$ of the discrete $\mathbb{Z}%
_{3}.$

\item[$(\mathbf{ii})$] \textbf{interval} $\mathrm{k}>3$: It concerns the
family of 2D lattices $W_{\mathrm{k}}^{\mathbf{su}_{3}}$ given by the
superposition $W_{\mathrm{k}}^{\mathbf{su}_{3}}=\mathbb{A}_{1}$ $\tbigcup $ $%
\cdots $ $\tbigcup $ $\mathbb{A}_{\mathrm{k}}$ where each sublattice\textrm{%
\ }$\mathbb{A}_{\mathrm{j}}$ is isomorphic to $R_{\mathrm{k}}^{\mathbf{su}%
_{3}}$. The weight lattice\textrm{\ }$W_{\mathrm{k}}^{\mathbf{su}_{3}}$ is
generated by rescaled weights\textrm{\ }$\mathbf{\tilde{\lambda}}^{i}=\sqrt{%
\frac{3}{\mathrm{k}}}\mathbf{\lambda }^{i}$ with intersection matrix $%
\mathcal{K}^{ij}=\frac{3}{\mathrm{k}}K^{ij}$. Similarly, the root lattices $%
R_{\mathrm{k}}^{\mathbf{su}_{3}}$ are generated by sites vectors $\mathbf{%
\tilde{\alpha}}_{i}=\sqrt{\frac{\mathrm{k}}{3}}\mathbf{\alpha }_{i}$ with $%
\mathcal{K}_{ij}=\frac{\mathrm{k}}{3}K_{ij}.$ We have%
\begin{equation}
\mathcal{K}_{ij}=\left(
\begin{array}{cc}
\frac{2\mathrm{k}}{3} & -\frac{\mathrm{k}}{3} \\
-\frac{\mathrm{k}}{3} & \frac{2\mathrm{k}}{3}%
\end{array}%
\right) \qquad ,\qquad \mathcal{K}^{ij}=\left(
\begin{array}{cc}
\frac{2}{\mathrm{k}} & \frac{1}{\mathrm{k}} \\
\frac{1}{\mathrm{k}} & \frac{2}{\mathrm{k}}%
\end{array}%
\right)
\end{equation}%
Here, the discriminant $W_{\mathrm{k}}^{\mathbf{su}_{3}}/R_{\mathrm{k}}^{%
\mathbf{su}_{3}}\simeq \mathbb{Z}_{\mathrm{k}}$ supports the k-fold
superposition (\ref{k}). The sublattices $\mathbb{A}_{\mathrm{j}}$ are
related to $\mathbb{A}_{\mathrm{1}}$ by global translations generated by $%
\mathbf{\tilde{\lambda}}_{1}$ and $\mathbf{\tilde{\lambda}}_{2}$, for
instance $\mathbb{A}_{\mathrm{j}}=\mathbb{A}_{\mathrm{j}}+(\mathrm{j}-1)%
\mathbf{\tilde{\lambda}}_{2}$.

\item[$(\mathbf{iii})$] \textbf{the range} $\mathrm{k}<3$ including k=1,2.%
\newline
The case k=1 is somehow peculiar, a subtle inversion occurs between the role
of $W_{\mathrm{1}}^{\mathbf{su}_{3}}$ and $R_{\mathrm{1}}^{\mathbf{su}_{3}}$
as evidenced by the unit cell areas $ucW_{\mathrm{1}}^{\mathbf{su}_{3}}=%
\sqrt{3}$ which has a larger area than $ucR_{\mathrm{1}}^{\mathbf{su}_{3}}=%
\frac{1}{\sqrt{3}}$ implying that $W_{\mathrm{2}}^{\mathbf{su}_{3}}\subset
R_{\mathrm{2}}^{\mathbf{su}_{3}}$. Interestingly$,$ for the irrational level
$\mathrm{k}=\sqrt{3}$, the unit cells\ coincide, implying $R_{\sqrt{3}}^{%
\mathbf{su}_{3}}\simeq W_{\sqrt{3}}^{\mathbf{su}_{3}}$. \newline
As for\textrm{\ }k=2, $W_{\mathrm{2}}^{\mathbf{su}_{3}}$ yields a 2D
honeycomb lattice with the superposition $\mathbb{A}$ $\tbigcup $ $\mathbb{B}%
.$ The inclusion\textrm{\ }$R_{\mathrm{2}}^{\mathbf{su}_{3}}\subset W_{%
\mathrm{2}}^{\mathbf{su}_{3}}$ still holds with unit cell areas satisfying%
\textrm{:} $ucR_{\mathrm{2}}^{\mathbf{su}_{3}}=\frac{4}{3}ucW_{\mathrm{2}}^{%
\mathbf{su}_{3}}.$ This configuration is depicted in Figure \textbf{\ref{35}}
where sites\textrm{\ }in $\mathbb{A}$ are colored red and those\textrm{\ }in
$\mathbb{B}$ are blue. However, there are no sites for $\mathbb{C}$.
\begin{figure}[tbph]
\begin{center}
\includegraphics[width=6cm]{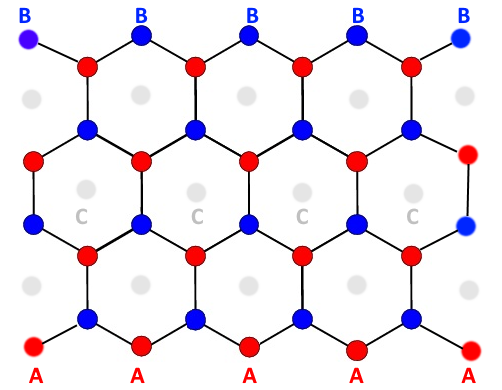}
\end{center}
\par
\vspace{-0.5cm}
\caption{The two sheets $\mathbb{A}$ $\tbigcup $ $\mathbb{B}$ of \textsc{W}$%
_{\mathrm{2}}^{\mathbf{su}_{3}}$. The sites in $\mathbb{A}$ are in red color
while the site B in $\mathbb{B}$ are in blue. The sites C of the \textsc{W}$%
_{\mathrm{3}}^{\mathbf{su}_{3}}$ are absent for k=2.}
\label{35}
\end{figure}
\end{description}

\subsection{Constructing four dimensional lattices}

Here, we use the 2D lattices $W_{\mathrm{k}}^{\mathbf{su}_{3}}$ and $R_{%
\mathrm{k}}^{\mathbf{su}_{3}}$\ described above to construct the 4D lattices
appearing in the inclusion relations $\mathbf{\Lambda }_{\mathrm{k}}^{%
\mathbf{su}_{3}}\mathbf{\subset \Lambda }_{\mathrm{k}\mathcal{C}}^{\mathbf{su%
}_{3}}\subset \mathbf{\Lambda }_{\mathrm{k}}^{\ast \mathbf{su}_{3}}$ \textrm{%
for} \textrm{k}$\geq 2$ where the largest lattice\textrm{\ }$\mathbf{\Lambda
}_{\mathrm{k}}^{\ast \mathbf{su}_{3}}$ is given by the cross product
\begin{equation}
\mathbf{\Lambda }_{\mathrm{k}}^{\ast \mathbf{su}_{3}}=W_{\mathrm{k}}^{%
\mathbf{su}_{3}}\times W_{\mathrm{k}}^{\prime \mathbf{su}_{3}}  \label{p}
\end{equation}%
with unit cell area $uc\mathbf{\Lambda }_{\mathrm{k}}^{\ast \mathbf{su}%
_{3}}=3/$\textrm{k}$^{2}<1$. The smallest even lattice $\mathbf{\Lambda }_{%
\mathrm{k}}^{\mathbf{su}_{3}}$ is realised as $\mathbf{\Lambda }_{\mathrm{k}%
}^{\mathbf{su}_{3}}=R_{\mathrm{k}}^{\mathbf{su}_{3}}\times R_{\mathrm{k}%
}^{\prime \mathbf{su}_{3}}$ whose unit cell area is given by $\mathrm{k}%
^{2}/3>1$. As for the intermediate even self dual $\mathbf{\Lambda }_{%
\mathrm{k}\mathcal{C}}^{\mathbf{su}_{3}},$ it can be realised as\textrm{\ }$%
R_{\mathrm{k}}^{\mathbf{su}_{3}}\times W_{\mathrm{k}}^{\prime \mathbf{su}%
_{3}}$ or equivalently $W_{\mathrm{k}}^{\mathbf{su}_{3}}\times R_{\mathrm{k}%
}^{\prime \mathbf{su}_{3}}$ with an area equal to 1. Recall that $W_{\mathrm{%
k}}^{\mathbf{su}_{3}}$ is given by the superposition $\mathbb{A}_{1}$ $%
\bigcup \mathbb{A}_{2}$ $\cdots $ $\bigcup $ $\mathbb{A}_{\mathrm{k}}$ with
the isomorphisms $\mathbb{A}_{\mathrm{1}}\simeq \cdots \simeq \mathbb{A}_{%
\mathrm{k}}\simeq R_{\mathrm{k}}^{\mathbf{su}_{3}}.$ Substituting into (\ref%
{p}), we find that $\mathbf{\Lambda }_{\mathrm{k}}^{\ast \mathbf{su}_{3}}$%
\textrm{\ }is defined as a superposition of k$^{2}$ 4D sublattices as\
follows%
\begin{equation}
\mathbf{\Lambda }_{\mathrm{k}}^{\ast \mathbf{su}_{3}}=(\bigcup\limits_{%
\mathrm{i=1}}^{\mathrm{k}}\mathbb{A}_{\mathrm{i}})\times (\bigcup\limits_{%
\mathrm{j=1}}^{\mathrm{k}}\mathbb{A}_{\mathrm{j}})=\bigcup\limits_{\mathrm{%
i,j=1}}^{\mathrm{k}}\mathbb{A}_{\mathrm{i}}\times \mathbb{A}_{\mathrm{j}}
\end{equation}%
where $\mathbb{A}_{\mathrm{i}}\times \mathbb{A}_{\mathrm{j}}\simeq \mathbf{%
\Lambda }_{\mathrm{k}}^{\mathbf{su}_{3}}.$ Regarding the realisation of the
even self dual lattice $\mathbf{\Lambda }_{\mathrm{k}\mathcal{C}}^{\mathbf{su%
}_{3}},$ there are k possible ways draw it within $\mathbf{\Lambda }_{%
\mathrm{k}}^{\ast \mathbf{su}_{3}}$:
\begin{equation}
\left( \mathbf{\Lambda }_{\mathrm{k}\mathcal{C}}^{\mathbf{su}_{3}}\right) _{%
\mathrm{i}}=\mathbb{A}_{\mathrm{i}}\times \bigcup\limits_{\mathrm{j=1}}^{%
\mathrm{k}}\mathbb{A}_{\mathrm{j}}
\end{equation}%
with unit cell area $uc\left( \mathbf{\Lambda }_{\mathrm{k}\mathcal{C}}^{%
\mathbf{su}_{3}}\right) _{\mathrm{i}}$ equals to $1.$ This relation shows
that the $\left( \mathbf{\Lambda }_{\mathrm{k}\mathcal{C}}^{\mathbf{su}%
_{3}}\right) _{\mathrm{i}}$'s form a k-cycle of $\mathbb{Z}_{\mathrm{k}}.$
Therefore, $\mathbf{\Lambda }_{\mathrm{k}}^{\ast \mathbf{su}_{3}}$ can be
also imagined as a superposition of the k self dual lattices $\left( \mathbf{%
\Lambda }_{\mathrm{k}\mathcal{C}}^{\mathbf{su}_{3}}\right) _{\mathrm{i}}$
namely
\begin{equation}
\mathbf{\Lambda }_{\mathrm{k}}^{\ast \mathbf{su}_{3}}=\bigcup\limits_{%
\mathrm{i=1}}^{\mathrm{k}}\left( \mathbf{\Lambda }_{\mathrm{k}\mathcal{C}}^{%
\mathbf{su}_{3}}\right) _{\mathrm{i}}\qquad \Leftrightarrow \qquad \mathbf{%
\Lambda }_{\mathrm{k}}^{\ast \mathbf{su}_{3}}=\left( \mathbf{\Lambda }_{%
\mathrm{k}\mathcal{C}}^{\mathbf{su}_{3}}\right) _{\mathrm{1}}\tbigcup \cdots
\tbigcup \left( \mathbf{\Lambda }_{\mathrm{k}\mathcal{C}}^{\mathbf{su}%
_{3}}\right) _{\mathrm{k}}
\end{equation}%
In what follows, we describe the lattices associated with the leading levels
k=2 and k=3.

\subsubsection{Case of k=2}

For this CS\ level, the $W_{\mathrm{2}}^{\mathbf{su}_{3}}$ is given by the
superposition of two sublattices like $\mathbb{A}\bigcup \mathbb{B}$ with
the isomorphisms $\mathbb{A}\simeq \mathbb{B}\simeq R_{\mathrm{2}}^{\mathbf{%
su}_{3}}.$ Substituting into (\ref{p}), we find that $\mathbf{\Lambda }_{%
\mathrm{2}}^{\ast \mathbf{su}_{3}}$ is constructed out of the superposition
of 4 sublattices like%
\begin{equation}
\mathbf{\Lambda }_{\mathrm{2}}^{\ast \mathbf{su}_{3}}=\left( \mathbb{A}%
\times \mathbb{A}\right) \text{ }\bigcup \text{ }\left( \mathbb{A}\times
\mathbb{B}\right) \text{ }\bigcup \text{ }\left( \mathbb{B}\times \mathbb{A}%
\right) \text{ }\bigcup \text{ }\left( \mathbb{B}\times \mathbb{B}\right)
\end{equation}%
Additionally, we have the property $\mathbf{\Lambda }_{\mathrm{2}}^{\ast
\mathbf{su}_{3}}=\left( \mathbf{\Lambda }_{\mathrm{k}\mathcal{C}}^{\mathbf{su%
}_{3}}\right) _{\mathrm{1}}$ $\bigcup $ $\left( \mathbf{\Lambda }_{\mathrm{k}%
\mathcal{C}}^{\mathbf{su}_{3}}\right) _{\mathrm{2}}$ with even self dual
sublattices $\left( \mathbf{\Lambda }_{\mathrm{k}\mathcal{C}}^{\mathbf{su}%
_{3}}\right) _{\mathrm{1}}$ and $\left( \mathbf{\Lambda }_{\mathrm{k}%
\mathcal{C}}^{\mathbf{su}_{3}}\right) _{\mathrm{2}}$ each made of two
sublattices as follows%
\begin{equation}
\begin{tabular}{lll}
$\left( \mathbf{\Lambda }_{\mathrm{k}\mathcal{C}}^{\mathbf{su}_{3}}\right) _{%
\mathrm{1}}$ & $=$ & $\mathbb{A}\times W_{\mathrm{2}}^{\mathbf{su}_{3}}$ \\
$\left( \mathbf{\Lambda }_{\mathrm{k}\mathcal{C}}^{\mathbf{su}_{3}}\right) _{%
\mathrm{2}}$ & $=$ & $\mathbb{B}\times W_{\mathrm{2}}^{\mathbf{su}_{3}}$%
\end{tabular}%
\qquad \Leftrightarrow \qquad
\begin{tabular}{lll}
$\left( \mathbf{\Lambda }_{\mathrm{k}\mathcal{C}}^{\mathbf{su}_{3}}\right) _{%
\mathrm{1}}$ & $=$ & $\left( \mathbb{A}\times \mathbb{A}\right) \text{ }%
\bigcup \text{ }\left( \mathbb{A}\times \mathbb{B}\right) $ \\
$\left( \mathbf{\Lambda }_{\mathrm{k}\mathcal{C}}^{\mathbf{su}_{3}}\right) _{%
\mathrm{2}}$ & $=$ & $\left( \mathbb{B}\times \mathbb{A}\right) \text{ }%
\bigcup \text{ }\left( \mathbb{B}\times \mathbb{B}\right) $%
\end{tabular}%
\end{equation}

\subsubsection{Case of k=3}

At k=3, the $W_{\mathrm{3}}^{\mathbf{su}_{3}}$ is given by the superposition
of three sublattices like $\mathbb{A}\bigcup \mathbb{B}\bigcup \mathbb{C}$
with isomorphisms $\mathbb{A}\simeq \mathbb{B}\simeq \mathbb{C}\simeq R_{%
\mathrm{3}}^{\mathbf{su}_{3}}.$ Accordingly\textrm{,} $\mathbf{\Lambda }_{%
\mathrm{3}}^{\ast \mathbf{su}_{3}}$ is given by the superposition of 9
isomorphic 4D sublattices:%
\begin{equation}
\begin{tabular}{lll}
$\mathbf{\Lambda }_{\mathrm{3}}^{\ast \mathbf{su}_{3}}$ & $=$ & $\left(
\mathbb{A}\times \mathbb{A}\right) \text{ }\bigcup \text{ }\left( \mathbb{A}%
\times \mathbb{B}\right) \text{ }\bigcup \text{ }\left( \mathbb{A}\times
\mathbb{C}\right) \text{ }\bigcup \text{ }$ \\
&  & $\left( \mathbb{B}\times \mathbb{A}\right) \text{ }\bigcup \text{ }%
\left( \mathbb{B}\times \mathbb{B}\right) \text{ }\bigcup \text{ }\left(
\mathbb{B}\times \mathbb{C}\right) \text{ }\bigcup \text{ }$ \\
&  & $\left( \mathbb{C}\times \mathbb{A}\right) \text{ }\bigcup \text{ }%
\left( \mathbb{C}\times \mathbb{B}\right) \text{ }\bigcup \text{ }\left(
\mathbb{C}\times \mathbb{C}\right) $%
\end{tabular}%
\end{equation}%
This decomposition can be recast as the union of three even self dual
lattices like $(\mathbf{\Lambda }_{\mathrm{3}\mathcal{C}}^{\mathbf{su}%
_{3}})_{1}$ $\tbigcup $ $(\mathbf{\Lambda }_{\mathrm{3}\mathcal{C}}^{\mathbf{%
su}_{3}})_{2}$ $\tbigcup $ $(\mathbf{\Lambda }_{\mathrm{3}\mathcal{C}}^{%
\mathbf{su}_{3}})_{3}$ with
\begin{equation}
\begin{tabular}{lll}
$(\mathbf{\Lambda }_{\mathrm{3}\mathcal{C}}^{\mathbf{su}_{3}})_{1}$ & $=$ & $%
\mathbb{A}\times W_{\mathrm{3}}^{\mathbf{su}_{3}}$ \\
$(\mathbf{\Lambda }_{\mathrm{3}\mathcal{C}}^{\mathbf{su}_{3}})_{2}$ & $=$ & $%
\mathbb{B}\times W_{\mathrm{3}}^{\mathbf{su}_{3}}$ \\
$(\mathbf{\Lambda }_{\mathrm{3}\mathcal{C}}^{\mathbf{su}_{3}})_{3}$ & $=$ & $%
\mathbb{C}\times W_{\mathrm{3}}^{\mathbf{su}_{3}}$%
\end{tabular}%
\end{equation}

We end this section by noticing that lattices with $4\times d$ dimensions
based on su(3) lattices can be obtained by taking tensor products, for
instance ($\mathbf{\Lambda }_{\mathrm{3}}^{\mathbf{su}_{3}})^{\otimes d}$
and ($\mathbf{\Lambda }_{\mathrm{3}}^{\ast \mathbf{su}_{3}})^{\otimes d}.$
Using $\mathbf{\Lambda }_{\mathrm{k}}^{\mathbf{su}_{3}}=(R_{\mathrm{k}}^{%
\mathbf{su}_{3}})^{\otimes 2}$ and $\mathbf{\Lambda }_{\mathrm{k}}^{\ast
\mathbf{su}_{3}}=(W_{\mathrm{k}}^{\mathbf{su}_{3}})^{\otimes 2},$ we get $%
(R_{\mathrm{k}}^{\mathbf{su}_{3}})^{\otimes 2d}$ and $(W_{\mathrm{k}}^{%
\mathbf{su}_{3}})^{\otimes 2d}$ while for the even self dual ($\mathbf{%
\Lambda }_{\mathrm{3}\mathcal{C}}^{\mathbf{su}_{3}})^{\otimes d},$ one
obtains $(R_{\mathrm{k}}^{\mathbf{su}_{3}})^{\otimes d}\times (W_{\mathrm{k}%
}^{\mathbf{su}_{3}})^{\otimes d}.$

\section{Conclusion and comments}

\begin{figure}[tbph]
\begin{center}
\includegraphics[width=8cm]{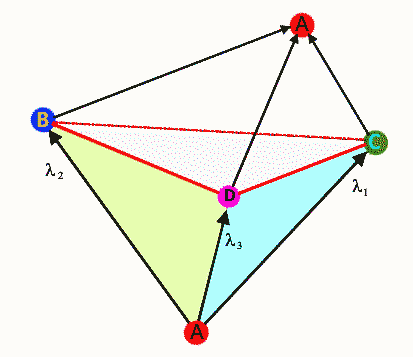}
\end{center}
\par
\vspace{-0.5cm}
\caption{three dimensional unit cell of the weight lattice of SU(4)
generated by $\mathbf{\protect\lambda }_{1},\mathbf{\protect\lambda }_{2},%
\mathbf{\protect\lambda }_{3}.$ The volume of this unit cell is equal to $%
uc(W^{\mathbf{su}_{3}})=1/2.$}
\label{u4}
\end{figure}

Narain CFTs form a class of two-dimensional CFTs that describe the geometry
of spacetime in string theory. They are characterized by vertex operators
whose left- and right- moving momenta of the form $\boldsymbol{p}_{\text{%
\textsc{l}/\textsc{r}}}\sim n^{i}\mathbf{\alpha }_{i}\pm m_{i}\mathbf{%
\lambda }^{i}$ spanning Narain lattices where the $\mathbf{\lambda }^{i}$'s
are dual to $\mathbf{\alpha }_{i}$'s. These lattices encode the
compactification data of the underlying string theory, specifically the
Kaluza-Klein modes $n^{i}\mathbf{\alpha }_{i}$ and windings $m_{i}\mathbf{%
\lambda }^{i}$. Narain theories are bosonic, non-chiral and modular
invariant when the corresponding lattices are even, Lorentzian and self-dual
$\mathbf{\Lambda }_{N\mathcal{C}}.$\textrm{\ }In the present investigation,
such lattices can be realised as the product $R^{\mathbf{su}_{N}}\times W^{%
\mathbf{su}_{N}}$ where\textrm{: }$\left( \mathbf{i}\right) $ the
Kaluza-Klein modes live in the su(N) root $R^{\mathbf{su}_{N}}$ lattice
generated by the $\mathbf{\alpha }_{i}$'s. We focus first on the N=2 case
describing standard code CFT and then explore N=3 as a non standard
extension. $\left( \mathbf{ii}\right) $ The winding modes occupy the weight $%
W^{\mathbf{su}_{N}}$ lattice generated by the dual $\mathbf{\lambda }_{i}$
with discriminant $W^{\mathbf{su}_{N}}/R^{\mathbf{su}_{N}}$ isomorphic to $%
\mathbb{Z}_{N}.$ In fact, this picture corresponds to the special model
where the CS level $\mathrm{k}=N.$ More generally, we have also studied
extensions to cover the intervals k\TEXTsymbol{>}N and k\TEXTsymbol{<}N for
both\textrm{\ }N=2 and N=3 cases. The same structure however extends
naturally to the full su(N) family which we briefly comment on below.\newline
The explicit construction presented in this paper for the su(2) and su(3)
models can be extended straightforwardly to higher rank Lie algebras su(N).
Starting from eq(\ref{plr}), that we rewrite like $\boldsymbol{h}%
=\sum_{i=1}^{N}n^{i}\mathbf{\alpha }_{i}$ and $\boldsymbol{q}%
=\sum_{i=1}^{N}m_{i}\mathbf{\lambda }^{i},$ we classify the possible
extensions according to\ the value for the CS level \textrm{k} including the
special case\textrm{\ k}$=N$ and \textrm{k}$>N.$ \newline
For $\mathbf{\alpha }_{i}.\mathbf{\alpha }_{j}=0,$ the lattices ($\mathbf{%
\Lambda }_{\mathrm{k}},\mathbf{\Lambda }_{\mathrm{k}\mathcal{C}},\mathbf{%
\Lambda }_{\mathrm{k}}^{\ast }$)\ with k=2 are embedded in the weight
lattice of
\begin{equation}
SU(2)^{d}\times SU(2)^{d}\simeq SO(4)^{d}
\end{equation}%
which is itself contained in $\mathbb{R}^{(d,d)}$ with signature (d,d). For
k=2 case, the root R$^{\mathbf{su}_{2}}$ and the weight W$^{\mathbf{su}_{2}}$
lattices are one dimensional with discriminant $W^{\mathbf{su}_{2}}/R^{%
\mathbf{su}_{2}}\simeq \mathbb{Z}_{\mathrm{2}}.$ When the level increases to
k\TEXTsymbol{>}2 , the discriminant generalises to $W_{\mathrm{k}}^{\mathbf{%
su}_{2}}/R_{\mathrm{k}}^{\mathbf{su}_{2}}\simeq \mathbb{Z}_{\mathrm{k}}$
indicating in turn that $W_{\mathrm{k}}^{\mathbf{su}_{2}}$ is constructed of
superposition of k isomorphic sublattices rather than just two sublattices $%
\mathbb{A}\tbigcup \mathbb{B}$ with the isomorphisms $\mathbb{A}\simeq
\mathbb{B}\simeq R^{\mathbf{su}_{2}}.$\newline
Considering non vanishing intersections $\mathbf{\alpha }_{i}.\mathbf{\alpha
}_{j}$ forming the entries of a Cartan matrix $K_{ij},$ one ends up with a
family of generalisations of the construction A of code CFT. These
extensions are\ naturally classified by finite dimensional Lie algebras
\textbf{g} with Cartan matrix $K_{ij}^{\mathbf{g}}.$ The corresponding three
lattices $\mathbf{\Lambda },$ $\mathbf{\Lambda }_{\mathcal{C}},$ $\mathbf{%
\Lambda }^{\ast }$ sit in the lattice of $\mathcal{G}^{d}\times \mathcal{G}%
^{d},$ where is the Lie group $\mathcal{G}^{d}$ associated with \textbf{g}$.$
In the SU(3) model investigated in section 3, the triplet ($\mathbf{\Lambda }%
_{\mathrm{k}},\mathbf{\Lambda }_{\mathrm{k}\mathcal{C}},\mathbf{\Lambda }_{%
\mathrm{k}}^{\ast }$)\ \textrm{for} $\mathrm{k}=3$ resides within the weight
lattice of $SU(3)^{d}\times SU(3)^{d}$\ contained in $\mathbb{R}^{(2d,2d)}$
with signature (2d,2d). For generic $\mathrm{k},$ we have the realisations
\begin{equation}
\mathbf{\Lambda }_{\mathrm{k}}\sim \left( R_{\mathrm{k}}^{\mathbf{su}%
_{3}}\right) ^{d}\times \left( R_{\mathrm{k}}^{\mathbf{su}_{3}}\right)
^{d}\qquad ,\qquad \mathbf{\Lambda }_{\mathrm{k}}^{\ast }\sim \left( W_{%
\mathrm{k}}^{\mathbf{su}_{3}}\right) ^{d}\times \left( W_{\mathrm{k}}^{%
\mathbf{su}_{3}}\right) ^{d}
\end{equation}%
with discriminant $W_{\mathrm{k}}^{\mathbf{su}_{3}}/R_{\mathrm{k}}^{\mathbf{%
su}_{3}}\simeq \mathbb{Z}_{\mathrm{k}}.$ Notice that for the special value
k=3, the 2D weight $W_{\mathrm{3}}^{\mathbf{su}_{3}}$ is given by the
triangular lattice while the 2D root $R_{\mathrm{3}}^{\mathbf{su}_{3}}$
forms its hexagonal sublattice. \newline
The construction built for $N=2,3$ extends directly the higher $SU(N)$ with
integer $N\geq 4$ \cite{SFB,SS}. For the $SU(4)$ class, with the special
value of the CS level k=4, the root $R^{\mathbf{su}_{4}}$ and the weight $W^{%
\mathbf{su}_{4}}$ are 4D lattices with discriminant $W^{\mathbf{su}_{4}}/R^{%
\mathbf{su}_{4}}\simeq \mathbb{Z}_{\mathrm{4}}$ indicating that $W^{\mathbf{%
su}_{4}}$ is made of the superposition of 4 lattices like
\begin{equation}
W^{\mathbf{su}_{4}}=\mathbb{A}\quad \tbigcup \quad \mathbb{B\quad \tbigcup
\quad C\quad \tbigcup \quad D}
\end{equation}%
where each of the 3D sublattices (with sites colored red, blue, green,
magenta) are isomorphic to the root lattice: $\mathbb{A}$ $\simeq $ $\mathbb{%
B\simeq C\simeq D}\simeq R^{\mathbf{su}_{4}}.$ The unit cell of the weight
lattice W$^{\mathbf{su}_{4}}$ is generated by the three fundamental weight
vectors $\left( \mathbf{\lambda }_{1},\mathbf{\lambda }_{2},\mathbf{\lambda }%
_{3}\right) ;$ it is depicted by the Figure \textbf{\ref{u4} }having four
colored sites\textbf{\ }extending the Figure \textbf{\ref{2i} }where only
three colored sublattices (red, blue, green) appear.
The volume $ucW^{\mathbf{su}_{4}}$ of the unit cell of the weight lattice is
given by the modulus $\left\vert \mathbf{\lambda }_{1}\wedge \mathbf{\lambda
}_{2}\wedge \mathbf{\lambda }_{3}\right\vert =\frac{1}{2}$; while the volume
$ucR^{\mathbf{su}_{4}}$ of the of unit cell of the root lattice, defined as $%
\left\vert \mathbf{\alpha }_{1}\wedge \mathbf{\alpha }_{2}\wedge \mathbf{%
\alpha }_{3}\right\vert ,$ is equal to $\sqrt{\det K_{\mathbf{su}_{4}}}=2$
with $K_{\mathbf{su}_{4}}$ being the Cartan matrix. Hence, the following
relation $ucR^{\mathbf{su}_{4}}=4\times ucW^{\mathbf{su}_{4}}$ indicating
that the unit cell of the root lattice R$^{\mathbf{su}_{4}}$ consists of
four unit cells of the weight W$^{\mathbf{su}_{4}}.$ Extending the
construction of sections 2 and 3, the 6D even self dual lattice $\mathbf{%
\Lambda }_{4\mathcal{C}}$ based on SU(4) is given for k=4 by the product $R^{%
\mathbf{su}_{4}}\times W^{\mathbf{su}_{4}}.$ It is contained into $\mathbf{%
\Lambda }_{4}^{\ast }=W^{\mathbf{su}_{4}}\times W^{\mathbf{su}_{4}}$ and it
contains $\mathbf{\Lambda }_{4}=R^{\mathbf{su}_{4}}\times R^{\mathbf{su}%
_{4}}.$

\section{Appendices:}

We give two appendices: the first concerns the properties of the self dual
lattices while the second covers the construction A$_{\mathbf{g}}$ of ref.
\cite{71}.

\subsection{Appendix A: self dual lattices}

A self-dual lattice $\mathbf{\Lambda }_{\mathcal{C}}$ is a lattice that
coincides with its dual lattice $\mathbf{\Lambda }_{\mathcal{C}}^{\ast },$
such that $\Lambda _{\mathcal{C}}=\Lambda _{\mathcal{C}}^{\ast }.$ Generally
speaking, a lattice $\mathbf{\Lambda }$ is a discrete subgroup of a
finite-dimensional real vector space $V=span\left\{ \mathbf{e}_{i}\right\} $
equipped with an inner product $\left\langle \ast ,\ast \right\rangle $; it
is generated by integer linear combinations of basis vectors of $V.$ The
dual lattice $\mathbf{\Lambda }^{\ast }$ is defined by
\begin{equation}
\mathbf{\Lambda }^{\ast }=\{\mathbf{x}\in V:\left\langle \mathbf{x},\mathbf{y%
}\right\rangle \in \mathbb{Z},\forall \mathbf{y}\in V\}
\end{equation}%
it is the set of all vectors $\mathbf{x}\in V$ such that the inner product $%
\left\langle \mathbf{x},\mathbf{y}\right\rangle $ is an integer for every $%
\mathbf{y}\in V.$ By setting $\mathbf{x}=\sum n^{i}\mathbf{e}_{i}$ and $%
\mathbf{y}=\sum m^{j}\mathbf{e}_{j}$ with integers $n^{i},m^{j}\in \mathbb{Z}%
,$ we have $\left\langle \mathbf{x},\mathbf{y}\right\rangle
=g_{ij}n^{i}m^{j} $ where the metric is $g_{ij}=\left\langle \mathbf{e}_{i},%
\mathbf{e}_{j}\right\rangle .$ The dual lattice can be viewed as the
"inverse" of the original lattice in terms of the bilinear form.\newline
Vectors $\mathbf{u}$ in $\mathbf{\Lambda }$ expand as $\sum_{i}u^{i}\mathbf{e%
}_{i}$ with components $u^{i}$ expressed in terms of the integers like $%
\sum_{j}\Lambda _{j}^{i}n^{j};$ as such they are characterised by the matrix
$\Lambda _{j}^{i}.$ Similarly, vectors $\mathbf{\tilde{u}}$ in $\mathbf{%
\Lambda }^{\ast }$ expand like $\sum \tilde{u}^{i}\mathbf{e}_{i}$ with $%
\tilde{u}^{i}=\sum_{j}\tilde{\Lambda}_{j}^{i}n^{j}$ and $\tilde{\Lambda}%
_{j}^{i}.$ A lattice $\mathbf{\Lambda }^{\ast }$ is self-dual if $\mathbf{%
\Lambda }=\mathbf{\Lambda }^{\ast }.$ Among the key properties of the self
dual $\mathbf{\Lambda }_{\mathcal{C}},$ we cite the following features:

\begin{itemize}
\item It is an integral lattice, meaning the inner product $\left\langle
\mathbf{x},\mathbf{y}\right\rangle $ of any two vectors in the lattice $%
\mathbf{\Lambda }_{\mathcal{C}}$ is an integer.

\item The characteristic matrix of $\mathbf{\Lambda }_{\mathcal{C}}$ has
determinant $\det \Lambda _{\mathcal{C}}=\pm 1;$ accordingly the volume of
its fundamental parallelepiped (primitive cell) is equal to 1.

\item Self-dual lattices are sometimes called \emph{unimodular lattices}
because their Gram matrix has determinant $\pm 1.$

\item Self-duality is preserved under orthogonal transformations\textrm{\ }%
(rotations and reflections).
\end{itemize}

Along with these features, the self dual lattices are important in many
areas such as coding theory, sphere packing, and modular forms \cite%
{pack,pack1}. Interesting examples of self-dual lattices include the
following: $\left( \mathbf{a}\right) $ the integer lattice $\mathbb{Z}^{n}$
in n-dimensional Euclidean space. $\left( \mathbf{b}\right) $ The $E_{8}$
lattice, having minimal norms equal to two, is the unique positive-definite
even unimodular lattice in 8 dimensions, making it a canonical example of a
self-dual lattice in that dimension \cite{E8}. It is generated by the simple
roots of the Lie algebra E$_{8}$. Its fundamental unit cell has volume 1,
which is equivalent to being equal to its dual lattice. $\left( \mathbf{c}%
\right) $ The Leech lattice \cite{Leech} which is a 24-dimensional even
unimodular lattice with no vectors of norm 2.

\subsection{Appendix B: construction A$_{\mathbf{g}}$}

The work \cite{71} gives a generalization of the relationship between
quantum error-correcting codes (QECC) and Narain conformal field theories
(NCFT) through lattice constructions. There, Narain CFTs correspond to code
lattices constructed from QECCs over the ring of cyclotomic field\emph{\
cyclotomic field} $\mathbb{Q}\left( \zeta _{p}\right) $ and its integer
subset $\mathfrak{O}\left( \zeta _{p}\right) $ defined as follows%
\begin{eqnarray}
\mathbb{Q}\left( \zeta _{p}\right) &=&\left\{
\tsum\limits_{i=0}^{p-2}a_{i}\zeta _{p}^{i}|\quad a_{i}\in \mathbb{Q},\text{
}\quad \text{\ }i=0,...,p-2\right\} \\
\mathfrak{O}\left( \zeta _{p}\right) &=&\left\{
\tsum\limits_{i=0}^{p-2}m_{i}\zeta _{p}^{i}|\quad m_{i}\in \mathbb{Z},\text{
}\quad \text{\ }i=0,...,p-2\right\}
\end{eqnarray}%
where $\zeta _{p}=e^{i\frac{2\pi }{p}\text{ }}$(the p-th root of unity with
prime $p\geq 3$) and the $a_{i}$ coefficients rational numbers. This
correspondence extends known constructions such as Construction A for binary
codes and its ternary analogues to a much broader class of codes over
cyclotomic integers. Recall that for construction A with binary codes, we
have
\begin{equation}
\Lambda _{\mathcal{C}}=\left\{ \frac{1}{\sqrt{\mathrm{k}}}\left( \boldsymbol{%
c}+\mathrm{k}\boldsymbol{m}\right) \quad |\quad \boldsymbol{c}\in \mathcal{C}%
,\quad \boldsymbol{m}\in \mathbb{Z}^{2n}\right\}
\end{equation}%
and the ternary homologue is given by%
\begin{eqnarray}
\Lambda _{\mathcal{C}}^{\mathbb{C}} &=&\left\{ \sqrt{\frac{2}{3}}\left[
\boldsymbol{c}\mathbf{+}\left( \omega -\bar{\omega}\right) \boldsymbol{m}%
\right] \quad |\quad \boldsymbol{c}\in \mathcal{C},\quad \boldsymbol{m}\in
\mathcal{E}^{2n}\right\} \\
\mathcal{E} &=&\left\{ m_{1}+m_{2}\omega \quad |\quad \omega =-\frac{1}{2}+i%
\frac{\sqrt{3}}{2}\right\}
\end{eqnarray}%
where $\omega =e^{i\frac{2\pi }{3}}$ and $\omega -\bar{\omega}=i\sqrt{3};$
and where $m_{1},m_{2}$\ are integers..\newline
Notice also that elements $\upsilon $ in the field $\mathbb{Q}\left( \zeta
_{p}\right) $ expand line $a_{0}+a_{1}\zeta _{p}+...+a_{p-2}\zeta _{p}^{p-2}$
with the trace property $tr\mathbf{\alpha }=\left( p-2\right)
a_{0}-a_{1}-...-a_{p-2}$ belonging to $\mathbb{Q}.$ Moreover, the pairing of
$x,y$ in $\mathfrak{O}\left( \zeta _{p}\right) $ is defined as $\left\langle
x,y\right\rangle =\frac{1}{p}tr\left( x\bar{y}\right) .$ The link with the
simple roots $\left\{ \alpha _{i}\right\} _{1\leq i\leq p-1}$ of su(p)
generating the root lattice $R^{\mathbf{su}_{p}}$ is given by
\begin{equation}
\alpha _{i}=\zeta _{p}^{i-1}-\zeta _{p}^{i},\qquad i=1,...,p-1
\end{equation}%
Similarly, for the\ fundamental weight vectors $\left\{ \lambda _{i}\right\}
_{1\leq i\leq p-1}$ of su(p) generating the root lattice $W^{\mathbf{su}%
_{p}},$ we have the following relations%
\begin{equation}
\lambda _{i}=1+\zeta _{p}+...+\zeta _{p}^{i-1},\qquad i=1,...,p-1
\end{equation}%
satisfying indeed $\left\langle \alpha _{i},\lambda _{j}\right\rangle
=\delta _{ij}.$ In this picture, we have:
\begin{equation}
\mathbf{\Lambda }_{\mathcal{C}}=\left\{ \mathbf{c}\lambda _{1}+\boldsymbol{m}%
\in \left( W^{\mathbf{su}_{p}}\right) ^{2n}|\quad c\in \mathcal{C},\quad
\vec{m}\in \left( R^{\mathbf{su}_{p}}\right) ^{2n}\right\}
\end{equation}%
The code-lattice construction A$_{\mathbf{g}}$ of \cite{71} is then a
reformulation of construction A for binary codes in terms of root R$^{%
\mathbf{g}}$ and weight W$^{\mathbf{g}}$ lattices of Lie algebras \textbf{g}%
, particularly those from the ADE classification a part the exceptional E$%
_{8}$ and $so(4k)$, enabling the construction of Narain lattices from codes
over more general rings $\mathbb{Z}_{q}$ with composite (non prime) integers
q. So, the construction A$_{\mathbf{g}}$ provides a unified framework
linking various QECCs over finite fields $\mathbb{F}_{\mathrm{p}}$ (with
prime p) or rings $\mathbb{Z}_{q}$ to Narain CFTs by embedding codes into
quotient rings of root R$^{\mathbf{g}}$ and weight W$^{\mathbf{g}}$ lattices
of Lie algebras. Overall, construction A$_{\mathbf{g}}$ generalizes and
unifies previous code/Narain CFT correspondences, revealing a rich interplay
between lattice theory, finite Lie algebras and conformal field theory. As
such, it broadens the scope of known correspondences and providing new tools
to explore both quantum information theory and string-theoretic CFTs.%
\begin{equation*}
\end{equation*}

\end{document}